\documentstyle[eqsecnum,aps,amstex,amssymb]{revtex}
\newcommand{\sigmabf}{\bbox{\sigma}}
\newcommand{\Sigmasf}{\bbox{\Sigma}}
\newcommand{\Sigmasfhat}{\bbox{\hat\Sigma}}
\newcommand{\nablabf}{\bbox{\nabla}}
\newcommand{\lambdabf}{\bbox{\lambda}}
\begin{document}
\makeatletter

\title{Functional Methods and Effective 
Potentials for Non-Linear Composites}

\author{Yves-Patrick Pellegrini\thanks{e-mail: {\tt pellegri@bruyeres.cea.fr}} and Marc Barth\'el\'emy\thanks{e-mail: {\tt mark@argento.bu.edu} until 08/31/99,  then {\tt barthele@bruyeres.cea.fr}}}
\address{Service de Physique de la Mati\`ere Condens\'ee, \\
Commissariat \`a l'\'Energie Atomique,\\
BP12, 91680 Bruyères-le-Châtel, France}
\author{Gilles Perrin\thanks{e-mail: {\tt gilles.perrin@ifp.fr}}}
\address{Institut Fran\c cais du Pétrole\\
1 et 4, avenue du Bois-Préau, 92852 Rueil-Malmaison Cedex, France.}
\date{First submitted: August 21, 1998. 
Revised version: May 3, 1999. Printed: \today}

\maketitle

\begin{abstract}
A formulation of variational principles in terms of functional 
integrals is proposed for any type of local plastic potentials. 
The minimization problem is reduced to the computation of a path 
integral. This integral can be used as a starting point for 
different approximations. As a first application, it is shown 
how to compute to second-order the weak-disorder perturbative 
expansion of the effective potentials in random composite. The 
three-dimensional results of Suquet and Ponte-Casta\~neda (1993) 
for the plastic dissipation potential with uniform applied 
tractions are retrieved and extended to any space dimension, 
taking correlations into account. In addition, the viscoplastic 
potential is also computed for uniform strain rates. 
\end{abstract}
\medskip

\medskip

{\em To be published in the Journal of Mechanics and Physics of Solids.}
\medskip

%-----------------------------------------------------------------------------
\section{Introduction}
In the last decade, studies of disordered non-linear composites 
have forced new homogeneization methods in Mechanics. Most such 
methods rely on rigorous variational principles, from which 
exact optimal bounds can be deduced (Ponte-Casta\~neda and 
Suquet, 1998). These principles are either standard minimum 
energy principles, or more refined approaches using a linear 
reference material, be it homogeneous (Hashin and Shtrikman, 
1962ab; Talbot and Willis, 1985), or even heterogeneous  
(Ponte-Casta\~neda, 1991; Suquet, 1993).

In the study of non-linear dielectric media as well, few exact 
results were previously available, save for the case where the 
non-linearity in the response was treated as a perturbation of 
the linear behavior (Stroud and Hui, 1988). 
In a recent pioneering work, Blumenfeld and Bergman (1989) 
computed the effective dielectric permittivity of a strongly 
non-linear random medium to second order in a weak-disorder 
expansion; cf.\ also Bergman and Lee (1998). 
The perturbative expansion was carried out by use of the Green 
function associated to the linear problem. 
An analogous approach was subsequently undertaken by Suquet and 
Ponte-Casta\~neda (1993) on strongly non-linear elastic composites. 
These results were then further extended to composites with 
inclusions of complex shapes (see, {\it eg.}, Ponte-Casta\~neda 
and Suquet (1998) for a review).

Perturbative results are important as testing-benches for 
self-consistent effective-medium theories. Linear self-consistent 
formulae (Budianski, 1965) are known to give qualitatively 
correct predictions even for a high contrast between the 
constituents of the random medium. They account for, e.g.\, 
the existence of a ``rigidity threshold'', i.e.\ a volumetric 
concentration of voids above which a porous linear elastic 
material loses its rigidity (Sahimi, 1998). In addition to the 
fulfillment of exact non-linear bounds (Gilormini, 1995), criteria 
of acceptability for a non-linear self-consistent formula include 
the recovery of second-order perturbative results, and also 
constraints on the critical behavior near the rigidity threshold 
(Levy and Bergman, 1994) where field fluctuations are enhanced. 
In Mechanics, a recent self-consistent effective-medium theory 
(Nebozhyn and Ponte-Casta\~neda, 1997) reproduces exact perturbative 
results to second order, but its critical fluctuations have not 
been studied yet. An equivalent theory has been proposed in 
electrostatics (Ponte-Casta\~neda and Kailasam, 1997). These 
result were derived from a variational principle.
 
As a complement to classical means, functional methods 
originating from field theory (for a book on functional methods 
see, e.g., Kleinert, 1995) and from the statistical physics of 
disordered systems (M\'ezard {\it et al.}, 1987) were recently 
harnessed to contribute to effective-medium studies (Barthélémy and 
Orland, 1993; Parcollet {\it et al.}, 1996). They have not yet spread 
in the mechanics community. Their major interest is that they not 
only provide a convenient workhorse for implementing variational 
principles under various constraints, but that they also lend 
themselves to approximations resulting in self-consistent 
effective-medium formulae (Budianski, 1965; Parcollet 
{\it et al.}, 1996), some of them being of a new type (Pellegrini 
and Barth\'el\'emy, 1999). With such methods, the problem of the 
minimization of the potential becomes equivalent to the computation 
of a functional integral, as shown below. This single integral, 
which may be written down for all types of local potentials, can 
then be approximated using standard tools (saddle point methods, 
perturbative expansions, etc.).

The purpose of this article is to introduce in detail the specific 
mathematical apparatus needed to apply functional methods to 
variational principles in mechanics. As a first application, 
we re-derive the result of Suquet and Ponte-Casta\~neda (1993) 
for the weak-disorder expansion of a strongly non-linear 
viscoplastic composite assumed to be of the Norton type. We 
extend it to any space dimension, and show how to take 
correlations into account.
 
We adopt an approach (Barth\'el\'emy and Orland, 1998) which 
utilizes the so-called ``replica method'' (Edwards and Anderson, 
1975) developed in the framework of spin glass theory (M\'ezard 
{\it et al.}, 1987).
The replica method enables one to average over the disorder in a 
non-perturbative way, but is by no means compulsory in order to 
exploit the functional formulation. It is however a most natural 
route towards extensions to self-consistent formulae. Specific 
applications to self-consistent formulae for linear or non-linear 
materials (including the EMT) will be presented elsewhere. 

The paper is self-contained, and is organized as follows. 
Section \ref{tpairf} introduces the variational formulations 
of the problem and explains how to cast them under a functional 
form, either starting from the local viscoplastic potential 
(expressed in terms of the stress tensor) or from the plastic 
dissipation potential (expressed in terms of the strain rate 
tensor). We show that minimizing a potential amounts to computing 
the statistical average of the logarithm of some suitably defined 
partition function. The computation is carried out with the 
replica method, explained in the text. In Sec.\ref{pc}, we 
specialize to the perturbative expansion, beginning with the 
plastic dissipation potential (Sec.\ \ref{gfpd}). The results 
are then applied to the Norton law (Sec.\ \ref{attnlpd}). We 
retrieve the result of Suquet and Ponte-Casta\~neda (1993), 
generalized to any space dimension.
The perturbative calculation of the viscoplastic potential, 
slightly more involved, is presented next, in Sec.\ \ref{tvp}. 
In both cases, the disorder is assumed to be {\em site-disorder}: 
the local potentials are statistically uncorrelated from point 
to point, for simplification purposes. This restriction is 
overcome in section \ref{otod}, where spatial correlations 
(not necessarily isotropic) are introduced.  Before concluding 
in Sec.\ \ref{co}, we comment on the range of applicability of 
the method (Sec.\ \ref{disc}).

Technical details are left to appendixes. Appendix \ref{ftim} 
is a brief reminder of the use of fourth-rank tensors in 
mechanics, and defines various fundamental fourth-rank tensors 
employed hereafter. Various algebras and sub-algebras encountered 
in the paper are also defined in this appendix, which ought be 
read before the reader goes through our calculations. The 
determinant ($\det$), trace ($\mathop{\text{tr}}$) and inversion 
$({}^{-1})$ operators will always be indexed by a label indicating 
in which algebra or sub-algebra of operators they act. Useful 
formulae for restricted gaussian integrals over vectors or second-rank 
tensors, which we could not find in the literature, are derived in 
Appendix \ref{gi}. Inversion and determinant formulae for matrices 
in the replica space are given in Appendix \ref{iadirs}. Evaluations 
of functional integrals are relegated to Appendix \ref{giotvasf}. 
The Legendre transform between the perturbative expansions of 
the viscoplastic and of the plastic dissipation potential is 
examined in Appendix \ref{tlt}.

Before going on, we state our notational conventions: unless 
otherwise indicated, scalars are denoted by regular typefaces 
(eg.\ $a$ or $A$); vectors are denoted by bold typefaces (${\bf a}$); 
tensors of rank two by sans-serif typefaces (${\sf a}$) [the exceptions are $\sigmabf$ and $\Sigmasf$ which appear in boldface], and tensors 
of rank four by capital Blackboard typefaces (${\mathbb A}$). Tensor 
indices are denoted by roman letters (e.g.\ $A_{ij}$), whereas 
replica indices are Greek letters. The Einstein summation convention 
on repeated indices is used, except when the indices are underlined 
($A_{ii}=\mathop{{\text tr}}{\sf A}$, whereas  $A_{\underline{i}\underline{i}}$ denotes 
the $i^{\text{th}}$ diagonal element of ${\sf A}$). Dyads (i.e.\ 
tensor products of vectors) are denoted omitting the tensor 
product operator: ${\sf A}={\bf a}{\bf b}$ is the dyad with 
components $A_{ij}=a_ib_j$.

%-----------------------------------------------------------------------------
\section{The problem and its formulation with replicas}
\label{tpairf}

%-----------------------------------------------------------------------------
\subsection{Variational principles and effective potentials}
We adopt here the presentation of Ponte-Casta\~neda and Suquet 
(1998). Let $\sigmabf({\bf x})$ be the local stress (tensor) 
field, and ${\sf d}({\bf x})$ be the local strain rate (tensor) 
field deriving from the local velocity (vector) field ${\bf v}
({\bf x})$. The material domain, of volume $V$, is denoted by 
$\Omega$. Then the equilibrium equations on $\sigmabf$ read, 
for ${\bf x}\in\Omega$ (the upper index ${}^{\text{t}}$ denotes 
the transpose):
\begin{eqnarray}
\nablabf.\sigmabf={\bf 0},\\
{\sigmabf}={}^{\text{t}}{\sigmabf}.
\end{eqnarray}
In addition, 
\begin{equation}
\label{dv}
{\sf d}={1\over 2}\left[\nablabf{\bf v}+{}^{\text{t}}
(\nablabf{\bf v})\right].
\end{equation}
The local constitutive relations are expressed by means of 
either the viscoplastic potential $\psi_{\bf x}$ or the plastic 
dissipation potential $\phi_{\bf x}$, as:
\begin{equation}
\sigmabf ={\partial\phi_{\bf x}\over\partial{\sf d}}
({\sf d}),\qquad {\sf d} ={\partial\psi_{\bf x}\over
\partial\sigmabf}(\sigmabf).
\end{equation}
The subscript ${}_{\bf x}$ indicates that the potentials may 
vary from point to point in the material, according to the 
local material properties. Both local potentials are usually 
convex functions of the fields, and are linked by the (Legendre) 
duality relation
\begin{equation}
\label{dual}
\phi_{\bf x}({\sf d})=\mathop{\text{max}}_{\sigmabf}
\left[\sigmabf\!:\!{\sf d}-\psi_{\bf x}({\sigmabf})\right].
\end{equation}

Classically, two types of boundary conditions are considered:

1) for boundary conditions of {\em uniform traction} (ut) 
on the boundary $\partial\Omega$ of the material, the 
heterogeneous material is shown to be macroscopically 
described by a pair of effective potentials $\Psi$ and 
$\Phi$ such that
\begin{mathletters} 
\label{unitract} 
\begin{eqnarray}
\label{unitract1} 
\Psi_{\text{ut}}({\Sigmasf})&=&\mathop{\text{min}}_{\sigmabf/
\left\{{\nablabf.\sigmabf={\bf 0}, \sigmabf={}^{\text{t}}\sigmabf
\atop \sigmabf.{\bf n}=\Sigmasf.{\bf n} \text{ on }\partial\Omega}
\right.}
\overline{\psi_{\bf x}(\sigmabf)},\\
\label{unitract2} 
\Phi_{\text{ut}}({\sf D})&=&\mathop{\text{min}}_{{\bf v}/
\overline{{\sf d}}={\sf D}}
\overline{\phi_{\bf x}({\sf d})}.
\end{eqnarray}
\end{mathletters} 
(the overline denotes a volume average on $\Omega$, and 
${\bf n}$ is the outward normal to $\Omega$).

2) for boundary conditions of {\em uniform strain rate} 
(us) ${\sf D}$ on the boundary, the effective potentials are:
\begin{mathletters}
\label{unistrain} 
\begin{eqnarray}
\label{unistrain1} 
\Psi_{\text{us}}({\Sigmasf})&=&
\mathop{\text{min}}_{\sigmabf/\left\{
{
\nablabf.\sigmabf={\bf 0}, {\sigmabf}=
{}^{\text{t}}{\sigmabf}
\atop
\overline{\sigmabf}={\Sigmasf}
}
\right.}
\overline{\psi_{\bf x}(\sigmabf)},\\
\label{unistrain2} 
\Phi_{\text{us}}({\sf D})&=&\mathop{\text{min}}_{{\bf v}/
{\bf v}={\sf D}.{\bf x} \text{ on }\partial\Omega}
\overline{\phi_{\bf x}({\sf d})}.
\end{eqnarray}
\end{mathletters}

For both types of boundary conditions, the effective 
potentials $\Phi$ and $\Psi$ are shown (Suquet, 1987; 
Willis, 1989) to be (Legendre) dual functions such that
\begin{equation}
\label{legendre}
\Psi(\Sigmasf)+\Phi({\sf D})=\Sigmasf\!:\!{\sf D},
\qquad {\sf D}={\partial \Psi(\Sigmasf)\over\partial
\Sigmasf},\qquad \Sigmasf={\partial \Phi({\sf D})\over
\partial{\sf D}}.
\end{equation}
In both cases $\Sigmasf=\overline{\sigmabf}$ and 
${\sf D}=\overline{\sf d}$. Depending on the boundary 
conditions considered, the latter equalities are either 
definitions, or theorems (Hill, 1963). Therefore only 
one equation in each pair (\ref{unitract}) or 
(\ref{unistrain}) is sufficient to characterize the 
effective homogeneized material.

The problem considered in this article consists in 
computing the effective potentials for disordered 
materials. Eqs.\ (\ref{unitract}), (\ref{unistrain}) 
refer to one sample (one particular realization of the 
disorder). A key assumption is that the effective 
potentials are {\em self-averaging}, i.e.\ that they 
do not depend on the sample at hand in the so-called 
``thermodynamic limit'', where the size of the sample goes 
to infinity. In other words, the material contains in 
the various regions of its bulk all the possible realizations 
of the disorder. Therefore, rather than carrying out the 
calculation of the effective potentials for one particular 
configuration of the disorder (which is impossible), we 
arrive at the correct result by considering statistical 
averages. For instance, in the limit $V\to\infty$ we can 
as well write, instead of (\ref{unitract1}):
\begin{equation}
\Psi_{\text{ut}}({\Sigmasf})=\left\langle\Psi_{\text{ut}}
({\Sigmasf})\right\rangle=\left\langle
\mathop{\text{min}}_{\sigma/\left\{{\nablabf.\sigmabf={\bf 0}, 
{\sigmabf}={}^{\text{t}}{\sigmabf}\atop \sigmabf.{\bf n}=\Sigmasf.{\bf n}
 \text{ on }\partial\Omega}\right.}
\overline{\psi_{\bf x}(\sigmabf)}\right\rangle.
\end{equation}
Note that the statistical averaging operator $\langle .\rangle$
and the infimum operator {\em do not commute} (inverting the 
operators would lead to the incorrect result that the effective 
potential is the average of the local potentials).

%-----------------------------------------------------------------------------
\subsection{Functional representations}
The minimization problem is reformulated by extending to 
functionals the following straightforward result for scalar functions:
\begin{equation}
\text{min}_{y\in [a,b]} f(y)=-\lim_{\beta\to\infty} 
{1\over\beta} \log \int_a^b dx\, e^{-\beta f(x)}.
\end{equation}
Note that this formula holds even if the minimum is met at 
several points in the interval.

Each infimum (\ref{unitract}) or (\ref{unistrain}) is obtained for 
the state of the fields $\sigmabf$ or ${\sf d}$ which minimize the 
potentials, among all the possible states fulfilling the equilibrium 
and boundary constraints. In statistical mechanics, this problem is 
analogous to that of finding the lowest energy state $E_0$ of some 
hamiltonian $H[s]$ functionally depending on some field $s({\bf x})$. 
In this case, $E_0$ is given by
\begin{mathletters} 
\begin{eqnarray}
\label{ezer}
E_0&=&-\lim_{\beta\to\infty}{1\over\beta}\log Z,\\
Z&=&\mathop{\text{tr}}\, e^{-\beta H},
\end{eqnarray}
\end{mathletters}
where the trace operator $\mathop{\text{tr}}$ here denotes the sum 
over all the allowed states of the system, i.e.\ the configurations 
of $s$. Indeed, the partition function $Z$ is obviously dominated 
by $\exp(-\beta E_0)$ in the limit $\beta\to\infty$, where $E_0\equiv H[s_0]$ 
and $s_0$ is the configuration of $s$ that minimizes $H$ ($\beta$ is 
the reciprocal of the temperature in statistical physics).

The means of summing over all the possible states of a continuous field 
subjected to constraints is to use functional integrals (also termed 
{\em path integrals}) with suitably defined measures. With such a 
formalism, it is more convenient to use the variational 
formulations (\ref{unitract2}) and (\ref{unistrain1}) where 
the boundary conditions are implemented through volume averages. 
We therefore write (with the additional statistical averages):

1) for uniform tractions [${\sf d}$ is computed in terms 
of ${\bf v}$ by (\ref{dv})],
\begin{equation}
\label{phiut}
\Phi_{\text{ut}}({\sf D})=-\lim_{\beta\to\infty}{1\over\beta}
\left\langle\log Z_{\text{ut}}\right\rangle,\qquad
Z_{\text{ut}}=\int {\cal D}\!{\bf v}\,\delta(\overline{\sf d}
-{\sf D})\,e^{-\beta\overline{\phi_{\bf x}({\sf d})}},
\end{equation}
where $\delta$ is the Dirac distribution and
\begin{mathletters}
\begin{eqnarray}
{\cal D}\!{\bf v}&=&\prod_{{\bf x}\in\Omega} d\!{\bf v}({\bf x}),\\
\delta(\overline{\sf d}-{\sf D})&=&\prod_{i\leq j}
\delta(\overline{ d}_{ij}- D_{ij});
\end{eqnarray}
\end{mathletters}
For brevity, the constrained measure in (\ref{phiut}) will 
be denoted by
\begin{equation}
{\cal D\!}_{\text{ut}}{\bf v}={\cal D}\!{\bf v}\,
\delta(\overline{\sf d}-{\sf D}).
\end{equation}

2) for uniform strain rates,
\begin{equation}
\label{psius}
\Psi_{\text{us}}(\Sigmasf)=-\lim_{\beta\to\infty}{1\over\beta}
\left\langle\log Z_{\text{us}}\right\rangle,\qquad
Z_{\text{us}}=\int{\cal D\!}_s \sigmabf\,\delta(\nablabf.\sigmabf)
\delta(\overline{\sigmabf}-\Sigmasf)\,e^{-\beta
\overline{\psi_{\bf x}(\sigmabf)}}.
\end{equation}
with
\begin{mathletters}
\begin{eqnarray}
{\cal D\!}_s \sigmabf&=&\prod_{{\bf x}\in\Omega} d_s 
\sigmabf({\bf x}),\\
d_s \sigmabf({\bf x})&=&2^{d(d-1)/4}[\prod_{i<j}\delta
\bigl(\sigma_{ij}({\bf x})-\sigma_{ji}({\bf x})\bigr)]
[\prod_{i,j}d\sigma_{ij}({\bf x})],\\
\delta(\nablabf.\sigmabf)&=&\prod_{i,{\bf x}\in\Omega}
\delta\bigl(\partial_k\sigma_{ki}({\bf x})\bigr),\\
\delta(\overline{\sigmabf}-\Sigmasf)&=&\prod_{i\leq j}
\delta(\overline{\sigma}_{ij}-\Sigma_{ij}).
\end{eqnarray}
\end{mathletters}
The integration measure $d_s \sigmabf({\bf x})$ allows 
one to integrate over all {\em symmetric} stress tensors 
(cf.\ Appendix \ref{gi}). The normalization factor 
$2^{d(d-1)/4}$ in $d_s \sigmabf({\bf x})$, where $d$ 
is the space dimension, is of no importance but has 
been included for consistency with Appendix \ref{gi}. 
We set:
\begin{equation}
{\cal D\!}_{\text{us}}\sigmabf={\cal D\!}_s \sigmabf\,
\delta(\nablabf.\sigmabf)\delta(\overline{\sigmabf}-\Sigmasf).
\end{equation}

%-----------------------------------------------------------------------------
\subsection{The replica method}
\label{trm}
Hereafter, we focus on $Z_{\text{ut}}$, but the following 
considerations apply {\em mutatis mutandis} to $Z_{\text{us}}$. 
The statistical average of $\log Z_{\text{ut}}$ is a complicated 
quantity to evaluate directly. The replica method consists 
in writing
\begin{equation}
\label{logz}
\left\langle\log Z_{\text{ut}}\right\rangle=\lim_{r\to 0} 
{1\over r}(\left\langle Z_{\text{ut}}^r\right\rangle-1),
\end{equation}
and in computing $\left\langle Z_{\text{ut}}^r\right\rangle$ 
with $r$ an integer (in the replica literature, the number of 
replicas, $r$, is usually denoted by $n$. We do not use the 
latter in order to avoid confusions with the Norton exponent). 
At the end of the calculation, we take the limit $r\rightarrow 0$, 
assuming the analytical continuation is legitimate, which we 
cannot prove in general. Therefore introducing $r$ 
{\em replicas} ${\bf v}^\alpha$ of the field ${\bf v}$, 
the partition function
\begin{equation}
Z_{\text{ut}}^r=\prod_{\alpha=1}^r\left(\int{\cal D\!}_{\text{ut}}
{\bf v}^\alpha\, e^{-\beta \overline{\phi_{\bf x}({\sf d}^\gamma)}}
 \right)=\int\prod_{\alpha=1}^r{\cal D\!}_{\text{ut}}{\bf v}^\alpha\, 
e^{-\beta \sum_{\gamma=1}^r\overline{\phi_{\bf x}({\sf d}^\gamma)}}
\end{equation}
is that of a system made of $r$ replicas of the initial system. The 
greek labels (which exclude $\beta$, this symbol being devoted to 
the reciprocal of the ``temperature'') are replica labels running 
from 1 to $r$. The statistical average therefore is
\begin{equation}
\label{avzed}
\left\langle Z_{\text{ut}}^r\right\rangle=\int\prod_{\alpha=1}^r
{\cal D\!}_{\text{ut}}{\bf v}^\alpha\, \left\langle e^{-\beta 
\sum_{\gamma=1}^r\overline{\phi_{\bf x}({\sf d}^\gamma)}}\right\rangle.
\end{equation}

For simplification purposes, we shall limit ourselves to {\em 
site-disordered} materials where the local potentials are 
statistically uncorrelated from point to point (we explain 
in Sec.\ \ref{otod} how to overcome this restriction). By 
definition of site disorder, we have for any function $F$ 
the property that \begin{equation}
\left\langle F\left(\psi_{\bf x}\bigl(\sigmabf({\bf x})\bigr)
\right)F\left(\psi_{\bf y}\bigl(\sigmabf({\bf y})\bigr)\right)
\right\rangle=\left\langle F\left(\psi_{\bf x}
\bigl(\sigmabf({\bf x})\bigr)\right)\right\rangle\left
\langle F\left(\psi_{\bf y}\bigl(\sigmabf({\bf y})\bigr)
\right)\right\rangle\text{ if }{\bf x}\not={\bf y}.
\end{equation}

For this type of disorder, the average in (\ref{avzed}) is 
readily carried out. Replacing the volume integral by a 
Riemann sum $\int d{\bf x}\to v\sum_{\bf x}$,   where $v$ 
denotes the unit ``infinitesimal'' volume element of the 
theory (here the size of $v$ also is the statistical 
correlation radius within which the potential is constant), 
one obtains 
\begin{eqnarray}
\left\langle Z_{\text{ut}}^r\right\rangle&=&\int
\prod_{\alpha=1}^r{\cal D\!}_{\text{ut}}{\bf v}^\alpha\, 
\left\langle e^{-(v\beta/ V) \sum_{{\bf x},{\gamma=1}}^r\phi_{\bf x}
({\sf d}^\gamma)}\right\rangle\nonumber\\
&=&\int\prod_{\alpha=1}^r{\cal D\!}_{\text{ut}}{\bf v}^\alpha\,
 \left\langle \prod_{\bf x} e^{-(v\beta/ V) \sum_{\gamma=1}^r\phi_{\bf x}
({\sf d}^\gamma)}\right\rangle\nonumber\\
&=&\int\prod_{\alpha=1}^r{\cal D\!}_{\text{ut}}
{\bf v}^\alpha\,\prod_{\bf x} \left\langle e^{-(v\beta/ V) 
\sum_{\gamma=1}^r\phi_{\bf x}({\sf d}^\gamma)}\right\rangle.
\end{eqnarray}
This finally yields:
\begin{equation}
\label{zrav}
\left\langle Z_{\text{ut}}^r\right\rangle=\int
\prod_{\alpha=1}^r{\cal D\!}_{\text{ut}}{\bf v}^\alpha\,
\exp \left[\int\! {d{\bf x}\over v}\log\left\langle 
e^{-(v\beta/ V) \sum_{\gamma=1}^r\phi_{\bf x}({\sf d}^\gamma)}
\right\rangle\right],
\end{equation}
where we reintroduce the integral notation in the exponent. This is 
the usual shorthand notation used in functional integral methods 
and has to be understood as the limit of a Riemann sum.

This expression, completed by (\ref{logz}) and (\ref{phiut}), 
constitutes an exact functional representation of the 
variational problem (\ref{unitract2}) for the particular 
type of disorder under study, and is valid for any form 
of $\phi_{\bf x}({\sf d})$. In other words, we recast a 
minimization problem under a computational form. Computing 
this integral exactly is equivalent to solving the minimization 
problem. We note that before averaging, the replicas are 
independent from one another, but are coupled after the 
average is carried out. Here appears an important feature 
of the replica method: the use of replicated fields allows 
one to average over disorder by transforming the disordered 
problem into a non-disordered one, but with couplings (if 
there were no couplings, averaging the logarithm would be 
a trivial matter). We are now left with a complicated 
field-theory problem without disorder, consisting of $r$ 
coupled fields, and we have to resort to approximations. 
In the next section, we present a perturbative calculation 
of this integral.

An important simplification arises from the fact that $\left
\langle Z_{\text{ut}}^r\right\rangle$ always shows up in the 
form $\left\langle Z_{\text{ut}}^r\right\rangle=(C\beta^c)^r 
f(r,\beta)$, where $f(0,x)=1$ and is a non-homogeneous 
function of $x$, $C$ is a $\beta$-independent constant, 
and $c$ is an exponent linked to $d$, the space dimension. 
Therefore
\begin{equation}
\label{simplif}
-{1\over\beta}{1\over r}(\left\langle Z_{\text{ut}}^r
\right\rangle-1)=-{1\over\beta}\left[\log(C\beta^c)+
{1\over r}\log f(r,\beta)+O(r)\right],
\end{equation}
and the multiplicative factor $(C\beta^c)^r$ does not 
contribute to the limit $\beta\to\infty$. Hereafter, such 
multiplicative factors will therefore systematically be 
dropped out of the calculations. This will be indicated 
by the proportionality symbol $\propto$ appearing instead 
of the equal sign in equations.

%-----------------------------------------------------------------------------
\section{Perturbative calculation}
\label{pc}
We detail in this section the perturbative calculation 
for $\Phi$ first, since it is simpler than that for $\Psi$. 
The latter is examined next, once the reader has become acquainted 
with the method. In the following, we assume that the $d$-dimensional 
material is rigid-viscoplastic, described by the dual [in the sense of 
(\ref{dual})] Norton potentials ($1\le n\le\infty$, $m=1/n$). 
\begin{equation}
\label{norton}
\left\{
\begin{array}{l}
\phi_{\bf x}({\sf d})=\theta_m
d_{\text{eq}}^{m+1}/(m+1)\\
\mathop{\text{tr}}_2({\sf d})=0
\end{array}
\right.
,\qquad \psi_{\bf x}(\sigmabf)=\omega_n\sigma_{\text{eq}}^{n+1}/(n+1),
\end{equation}
 with
\begin{equation}
\theta_m=\sigma_0/\dot{\varepsilon}^m_0,\qquad 
\omega_n=\dot{\varepsilon}_0/\sigma_0^n.
\end{equation}
The equivalent Mises norms $\sigma_{\text{eq}}$ 
and $d_{\text{eq}}$ are:
\begin{equation}
d_{\text{eq}}=\left[{d-1\over d}\mathop{\text{tr}}_2
({\sf d}^{\prime 2})\right]^{1/2},\qquad \sigma_{\text{eq}}
=\left[{d\over d-1}\mathop{\text{tr}}_2({\sigmabf}^{\prime 2})\right]^{1/2}.
\end{equation}
These are the $d$-dimensional counterparts of the usual 
three-dimensional definitions (for $d=3$, we recover the 
factors $2/3$ and $3/2$). The deviatoric part of the stress 
or strain rate tensor has been denoted by a prime: ${\sf a}'
={\sf a}-\mathop{\text{tr}}_2({\sf a}){\sf I}/d$, where 
${\sf a}=\sigmabf$ or ${\sf d}$.

The additional incompressibility constraint 
$\mathop{\text{tr}}_2({\sf d})=\nablabf.{\bf v}=0$, 
which characterize incompressible materials, can as well be 
accounted for by adding a term $(K/2)\mathop{\text{tr}}_2({\sf d})^2$ 
to $\phi_{\bf x}({\sf d})$, and by letting the constant $K\to\infty$
 at the end of the calculation. Instead, we shall implement it 
directly by multiplying the measure ${\cal D}_{\text{ut}}
{\bf v}$ by a factor $\delta(\nablabf.{\bf v})$. 
Both ways actually amount to the same.

The constitutive parameters $\sigma_0$ and $\dot{\varepsilon_0}$ 
decribe the material and are random functions of the position 
variable ${\bf x}$, uncorrelated from point to point, according 
to our site-disorder hypothesis. The exponent $n$ is taken to 
be constant in the material, for simplification purposes (but 
this is not a limitation of the method).

%-----------------------------------------------------------------------------
\subsection{The plastic dissipation potential}
\label{tpdp}

%-----------------------------------------------------------------------------
\subsubsection{General framework}
\label{gfpd}
Integral (\ref{zrav}) is an exact expression for the effective 
potential. Its perturbation expansion is considered hereafter. 
The measure ${\cal D}_{ut}{\bf v}^{\alpha}$ contains the term 
$\delta(\overline{\sf d}-{\sf D})$. This means that the main 
contribution to this integral is coming from fields around ${\sf D}$. 
Indeed, if we suppose that only the constant field  ${\sf d}={\sf D}$ 
(i.e.\ ${\bf v}={\sf D}.{\bf x}$ such that $\mathop{\text{tr}}_2({\sf D})=0$)
 contributes to the integral, we find the trivial lowest order result
\begin{equation}
\Phi_{\text{ut}}({\sf D})=\langle\phi_{\bf x}({\sf D})\rangle.
\end{equation}
We shall now seek the first corrective term, and compute (\ref{zrav}) 
by integrating over fields ${\bf v}$ such that ${\sf d}$ differs 
little from ${\sf D}$. The perturbation expansion is obtained with 
Laplace's method for asymptotic expansions of integrals (Bender 
and Orszag, 1978). We therefore write
\begin{equation}
{\sf d}={\sf D}+{\sf e}
\end{equation}
and expand the logarithm in (\ref{zrav}) in powers of ${\sf e}$. The linear 
and quadratic terms are then kept inside the main exponential, and the 
rest is further expanded in ${\sf e}$. We start from the expansion:
\begin{equation}
\label{phiexpan}
\phi_{\bf x}({\sf D}+{\sf e})=\phi_{\bf x}({\sf D})+W'_{ij}e_{ji}+{1\over 2}
 W''_{ij,kl}e_{ji}e_{lk}+O({\sf e}^3),
\end{equation}
with
\begin{equation}
\label{derivphi}
W'_{ij}({\bf x})=\left.{\partial\phi_{\bf x}\over\partial d_{ij}({\bf x})}
\right|_{{\sf d}={\sf D}},\qquad W''_{ij,kl}({\bf x})=\left.{\partial^2
\phi_{\bf x}\over\partial d_{ij}({\bf x})\partial d_{kl}({\bf x})}
\right|_{{\sf d}={\sf D}}.
\end{equation}
With a change of integration variables such that ${\bf v}$ now denotes 
the correction to ${\sf D}.{\bf x}$, the expansion of $\left\langle 
Z_{\text{ut}}^r\right\rangle$ takes the form (taking into account 
the incompressibility constraint):
\begin{eqnarray}
\label{zutexpans}
&&\left\langle Z_{\text{ut}}^r\right\rangle=\langle e^{-r\beta 
{v\over V} \phi_{\bf x}({\sf D})}
\rangle^{V\over v}
\int \prod_{\alpha}{\cal D\!}_{\text{ut}}{\bf v}^\alpha
\delta(\nablabf.{\bf v}^\alpha)\left[1+O({\sf e}^3)\right]
\nonumber\\ 
\label{sms}
&&{}\times\exp\left[-\beta\langle {\sf W}'\rangle_e\!:\!\sum_{\alpha}
 \overline{\sf e}^\alpha-{1\over 2}{\beta\over V}\sum_{\alpha\gamma}
\int\!d{\bf x}\,{\sf e}^\alpha({\bf x})\!:\!{\mathbb M}^{\alpha\gamma}
\!:\!{\sf e}^\gamma({\bf x})\right].
\end{eqnarray}
The functional measure now is ${\cal D\!}_{\text{ut}}{\bf v}={\cal D\!}
{\bf v}\,\delta(\overline{\sf e})$, and ${\sf e}=[\nablabf{\bf v}+
{}^{\text{t}}(\nablabf{\bf v})]/2$. Besides, we have introduced 
the matrix ${\mathbb M}^{\alpha\gamma}={\mathbb M}_1\delta_{\alpha\gamma}
-\beta(v/V){\mathbb M}_2$, with
\begin{mathletters}
\label{mwphi}
\begin{eqnarray}
{\mathbb M}_1&=&\langle {\mathbb W}''\rangle_e\,\\
{\mathbb M}_2&=&\langle {\sf W}'{\sf W}'\rangle_e-\langle 
{\sf W}'\rangle_e\langle {\sf W}'\rangle_e.
\end{eqnarray}
\end{mathletters}
The notation $\langle\cdot\rangle_e$ stands for the weighted average
\begin{equation}
\label{ave}
\langle A({\bf x})\rangle_e\equiv {
\langle A({\bf x})
e^{-r\beta {v\over V} \phi_{\bf x}({\sf D})}\rangle
\over
\langle e^{-r\beta {v\over V} \phi_{\bf x}({\sf D})}
\rangle}=\langle A({\bf x})\rangle+O(r).
\end{equation}
Only the leading term in $r$ is needed in (\ref{ave}) [a handwaving 
argument for this is that the sums over replicas in (\ref{sms}) are 
at least proportional to $r$: the contribution is null for $r=0$. 
They therefore already provide the linear terms in $r$ required for 
evaluating (\ref{logz})], and we shall therefore in practice identify 
$\langle .\rangle_e$ with $\langle .\rangle$. Hereafter, square matrices 
in the replica space are denoted by a double overbar. The identity in 
the replica space is $\smash{I^{\alpha\gamma}=\delta_{\alpha\gamma}}$, 
and we furthermore define the matrix $\smash{U^{\alpha\gamma}=1}$ for 
all $\alpha,\gamma$, so that
\begin{equation}
\label{mphi}
\overline{\overline{\mathbb M}}={\mathbb M}_1\overline{\overline{I}}
-\beta(v/V){\mathbb M}_2\overline{\overline{U}}.
\end{equation}

The second-order term in the weak-disorder expansion of $\left\langle 
Z_{\text{ut}}^r\right\rangle$ is provided by the gaussian integral 
(\ref{sms}) only, neglecting the corrections $\smash{O({\sf e}^3)}$ 
and higher. As in centered scalar gaussian integrals, the term proportional 
to $\smash{{\sf e}^3}$ vanishes upon integration, and the next non-zero 
correction in fact is an $\smash{O({\sf e}^4)}$. The aim of the paper 
being to explain the functional method, we shall ignore such higher-order 
terms hereafter but they could as well be computed by pursuing 
Laplace's expansion scheme.

The details of the integration over ${\bf v}$ with measure 
${\cal D\!}_{\text{ut}}{\bf v}$ are left to Appendix \ref{giotvasf}. 
Taking into account that ${\sf e}$ is expressed in terms of the 
gradient of ${\bf v}$ involves a change from ${\bf v}({\bf x})$ to 
its Fourier components ${\bf v}({\bf k})$. One finds
\begin{equation}
\label{zutnav}
\left\langle Z_{\text{ut}}^r\right\rangle\propto\langle 
e^{-r\beta {v\over V} \phi_{\bf x}({\sf D})}
\rangle^{V\over v}\prod_{\bf k\not={\bf 0}}\left\{\det_{2[\perp{\bf\hat k}],
\text{rep}}\left[{\sf Q}_{\bf\hat k}.\left({\bf\hat k}.
\overline{\overline{{\mathbb M}}}.{\bf\hat k}\right).{\sf Q}_{\bf\hat k}\right]
\right\}^{-1/2},
\end{equation}
where ${\bf\hat k}={\bf k}/k$ and ${\sf Q}_{\bf\hat k}=
{\sf I}-{\bf\hat k}{\bf\hat k}$. 

We now calculate the determinant. Defining
\begin{mathletters}
\begin{eqnarray}
{\sf M}_1&=&{\sf Q}_{\bf\hat k}.({\bf\hat k}.{\mathbb M}_1.
{\bf\hat k}).{\sf Q}_{\bf\hat k},\\
{\sf M}_2&=&{\sf Q}_{\bf\hat k}.({\bf k}.{\mathbb M}_2.{\bf k}).
{\sf Q}_{\bf\hat k},
\end{eqnarray} 
\end{mathletters}
we obtain, with the help of formulae (\ref{invdet}) in Appendix 
\ref{iadirs}:
\begin{equation}
\det_{2[\perp{\bf\hat k}],\text{rep}}\left[{\sf Q}_{\bf\hat k}.
\left({\bf\hat k}.\overline{\overline{{\mathbb M}}}.
{\bf\hat k}\right).{\sf Q}_{\bf\hat k}\right]=1+\left[
\log\det_{2[\perp{\bf\hat k}]}({\sf M}_1)-\beta{v\over V}
\mathop{\text{tr}}_{2}\left({\sf M}_2.{\sf M}_1^{-1_{2[
\perp{\bf\hat k}]}}\right)\right]r+O(r^2).
\end{equation}
Carrying out as in (\ref{logz}) the limit $r\to 0$ in 
(\ref{zutnav}) it follows that
\begin{eqnarray}
-{1\over\beta}\langle \log Z_{\text{ut}}\rangle&=&\langle
\phi_{\bf x}({\sf D})\rangle-{1\over 2}{v\over V}
\sum_{\bf k\not={\bf 0}}\mathop{\text{tr}}_{2}\left({\sf M}_2.
{\sf M}_1^{-1_{2[\perp{\bf\hat k}]}}\right)\nonumber\\
\label{logzut}
&&{}+o(\log(\beta)/\beta)+o(1/\beta).
\end{eqnarray}
The sum over ${\bf k\not={\bf 0}}\ldots$ (divided by $V$) can 
now be replaced by a $d$-dimensional integral [divided by $(2\pi)^d$] 
over all the Fourier modes ${\bf k}$ in the continuum limit, since 
the integrand is not singular at $k=0$. The finite size of the 
elementary cells of volume $v$ is accounted for by an upper 
cut-off $\Lambda$ in this integral. It is chosen such that
\begin{equation}
v\int_{k<\Lambda} {d{\bf k}\over(2\pi)^d}= 1.
\end{equation}
Hence $\Lambda\to\infty$ when $v\to 0$. We denote by 
$S_d=2\pi^{d/2}/\Gamma(d/2)$ the area of the sphere of unit 
radius in ${\bf R}^d$, and by $d\Omega_{\bf\hat k}$ the 
integration measure over the solid angle in the direction 
${\bf\hat k}$ ($d{\bf k}=k^{d-1}\,dk\,d\Omega_{\bf\hat k}$).

Letting $\beta\to\infty$ in (\ref{logzut}) we deduce, by 
(\ref{phiut}): 
\begin{equation}
\label{phiutdsf}
\Phi_{\text{ut}}({\sf D})=\langle\phi_{\bf x}({\sf D})\rangle
-{1\over 2}\int{d\Omega_{\bf\hat k}\over S_{d}}
\mathop{\text{tr}}_{2}\left({\sf M}_2.{\sf M}_1^{-1_{2[\perp{\bf\hat k}]}}\right).
\end{equation}
This result is the second-order perturbation expansion 
valid for any potential $\phi_{\bf x}({\sf d})$ with the 
constraint $\mathop{\text{tr}}_2 {\sf d}=0$.

%-----------------------------------------------------------------------------
\subsubsection{Application to the Norton law}
\label{attnlpd}
We evaluate the previous result (\ref{phiutdsf}) for the 
Norton potential (\ref{norton}). We write
\begin{equation}
\theta_m=\langle \theta_m\rangle+\delta\theta_m.
\end{equation}
Defining ${\sf\hat D}={\sf D}'/\sqrt{\mathop{\text{tr}}_2
({\sf D}^{'2})}$, so that $\mathop{\text{tr}}_{2}({\sf\hat D}.
{\sf\hat D})=\mathop{\text{tr}}_{4}({\sf\hat D}{\sf\hat D})=1$, 
(\ref{derivphi}) and (\ref{mwphi}) entail:
\begin{mathletters}
\label{MMphi}
\begin{eqnarray}
{\mathbb M}_1&=&m_1 {\mathbb J}+m_2{\sf\hat D}{\sf\hat D},\\
{\mathbb M}_2&=&m_3{\sf\hat D}{\sf\hat D},
\end{eqnarray}
\end{mathletters}
where
\begin{mathletters}
\label{mmmphi}
\begin{eqnarray}
m_1&=&{d-1\over d}\langle\theta_m\rangle D_{\text{eq}}^{m-1},\\
m_2&=&(m-1)m_1,\\
m_3&=&{d-1\over d}\left\langle\delta\theta_m^2
\right\rangle D_{\text{eq}}^{2m}.
\end{eqnarray}
\end{mathletters}
Setting
\begin{equation}
\label{dhatk}
D_{\bf\hat k}=|{\sf Q}_{\bf\hat k}.{\sf\hat D}.{\bf\hat k}|^2
={\bf\hat k}.{\sf\hat D}^2.{\bf\hat k}-({\bf\hat k}.{\sf\hat D}
.{\bf\hat k})^2,
\end{equation}
and introducing the unit vector ${\bf\hat u}={\sf Q}_{\bf\hat k}
.{\sf\hat D}.{\bf\hat k}/|{\sf Q}_{\bf\hat k}.{\sf\hat D}.
{\bf\hat k}|$,
we obtain
\begin{mathletters}
\begin{eqnarray}
{\sf M}_1&=&{m_1\over 2}\left({\sf Q}_{\bf\hat k}-{\bf\hat u}
{\bf\hat u}\right)+\left(m_2 D_{\bf\hat k} +{m_1\over 2}\right)
{\bf\hat u}{\bf\hat u},\\
{\sf M}_2&=&m_3 D_{\bf\hat k}\,{\bf\hat u}{\bf\hat u}.
\end{eqnarray}
\end{mathletters}
Since ${\sf Q}_{\bf\hat k}-{\bf\hat u}{\bf\hat u}$ and ${\bf\hat u}
{\bf\hat u}$, are a pair of complementary orthogonal projectors 
in $L(2[\perp{\bf\hat k}])$ (where ${\sf Q}_{\bf\hat k}$ plays 
the role of the identity), 
the inverse of ${\sf M}_1$, its product with ${\sf M}_2$ and 
the final trace in (\ref{phiutdsf}) are readily obtained, resulting in:
\begin{equation}
\label{simplephi}
\Phi_{\text{ut}}({\sf D})=\langle\phi_{\bf x}({\sf D})\rangle
-\int{d\Omega_{\bf\hat k}\over S_{d}}{m_3 D_{\bf\hat k}
\over m_1+2 m_2 D_{\bf\hat k}}.
\end{equation}
Now replacing $m_1$, $m_2$ and $m_3$ by their values 
(\ref{mmmphi}) finally yields
\begin{mathletters}
\label{nortplast}
\begin{eqnarray}
\Phi_{\text{ut}}({\sf D})&=&{\theta_{m\,\text{eff}}
({\sf D})\over m+1}D_{\text{eq}}^{m+1},\\
\label{thetaeff}
\theta_{m\,\text{eff}}({\sf D})&=&\langle\theta_m\rangle
\left[1-{\left\langle\delta\theta_m^2\right\rangle\over 
\langle\theta_m\rangle^2}\int{d\Omega_{\bf\hat k}\over 
S_{d}}{(m+1)D_{\bf\hat k}\over 1+2(m-1)D_{\bf\hat k}}
+O(\left\langle\delta\theta_m^3\right\rangle/\left\langle
\theta_m\right\rangle^3)\right].
\end{eqnarray}
\end{mathletters}
This result reduces to that of Suquet and Ponte-Casta\~neda 
(1993) for a space dimension $d=3$.

%-----------------------------------------------------------------------------
\subsection{The viscoplastic potential}
\label{tvp}
We undertake a similar calculation for the viscoplastic potential. 
It can be checked (cf.\ Appendix E) that the resulting perturbative 
expression  (\ref{nortvis}) is linked by a Legendre transform to 
(\ref{nortplast}), the one for the dissipation potential, though 
different boundary conditions are considered. This is a consequence 
of our working in the thermodynamic limit where the volume of the 
system $V\to\infty$ (in the calculations, this shows up essentially 
in that the Fourier sums are carried out with $k>0$, rather than 
with $k>(2\pi)/L$, where $L$ would be the typical system size). As 
noticed by Hill and Mandel in elasticity, the homogeneized stiffness 
tensor obtained with uniform strains on $\partial\Omega$, and the 
homogeneized compliances tensor obtained with uniform stresses on 
$\partial\Omega$ are inverse of one another up to terms of order 
$O\bigl( (\xi/L)^3\bigr)$ in three dimensions, where $\xi$ is the 
typical size of the heterogeneities \cite{SUQU87}. They therefore 
become exact inverses in the thermodynamic limit. The same effect 
is at play here. The viscoplastic potential can thus directly be 
obtained from the dissipation potential as done in Appendix E. However 
we prefer to compute it from scratch because differences with the 
previous section show up, and it may be useful to see what these are 
for future purposes: indeed, as soon as one leaves exact perturbative 
expansions and deals with self-consistent estimates, one approximation 
scheme worked out separately for dual potentials quite often generates 
two estimates which are not duals of one another (a notable exception 
is the self-dual EMT theory). Treating in a symmetrical way the two 
potentials is then a more sensible thing to do than working with only 
one.

%-----------------------------------------------------------------------------
\subsubsection{General framework}
\label{gfvp}
We now have to compute
\begin{equation}
\label{zravus}
\left\langle Z_{\text{us}}^r\right\rangle=\int\prod_{\alpha=1}^r
{\cal D\!}_{\text{ut}}{\sf \sigma}^\alpha\,\exp \left[\frac{1}{v}
\int d{\bf x}\log\left\langle e^{-\beta(v/V) \sum_{\gamma=1}^r
\psi_{\bf x}({\sf \sigma}^\gamma)}\right\rangle\right].
\end{equation}
The perturbative expansion of the effective potential is carried 
out around the constant solution $\sigmabf=\Sigmasf$, writing 
$\sigmabf=\Sigmasf+{\sf s}$. The expansion of $\psi_{\bf x}
(\Sigmasf+{\sf s})$ has the same form as (\ref{phiexpan}). We 
now denote by ${\sf W}'$ and ${\mathbb W}$ the first and second 
derivatives of $\psi_{\bf x}(\sigmabf)$ evaluated at 
$\sigmabf=\Sigmasf$, respectively. Using ${\sf s}$ as integration 
variables, the expansion of $\left\langle Z_{\text{us}}^r\right\rangle$ is:
\begin{eqnarray}
&&\left\langle Z_{\text{us}}^r\right\rangle\propto\langle 
e^{-r\beta {v\over V} \psi_{\bf x}(\Sigmasf)}
\rangle^{V\over v}\int \prod_{\alpha}\tilde{\cal D\!}_s
{\sf s}^\alpha\left[1+O({\sf s}^3)\right]\nonumber\\
\label{sms2}
&&{}\times 
\exp\left[-\beta{v\over V}\langle {\sf W}'\rangle_e\!:\!\sum_{\alpha}
 \overline{\sf s}^\alpha-{1\over 2}\beta{1\over V}\sum_{\alpha\gamma}
\int d{\bf x}\,{\sf s}^\alpha({\bf x})\!:\!{\mathbb M}^{\alpha\gamma}
\!:\!{\sf s}^\gamma({\bf x})\right],
\end{eqnarray}
where $\tilde{\cal D\!}_s{\sf s}={\cal D\!}_s{\sf s}\,\delta(\nablabf.
{\sf s})\delta(\overline{\sf s})$ and ${\mathbb M}^{\alpha\gamma}$ 
is defined as in Eqs.\ (\ref{mwphi}) and (\ref{mphi}). Once again, 
the details of the integration over ${\sf s}$ with measure $\tilde{\cal D\!}_s
{\sf s}$ are left to Appendix \ref{giotvasf}. The implementation of the
 constraint $\nablabf.{\sf s}=0$ involves a change from ${\sf s}({\bf x})$ 
to its Fourier components ${\sf s}({\bf k})$. The result is:
\begin{equation}
\label{znav}
\left\langle Z_{\text{us}}^r\right\rangle\propto\langle 
e^{-r\beta {v\over V} \psi_{\bf x}(\Sigmasf)}
\rangle^{V\over v}\prod_{\bf k\not={\bf 0}}\left\{
\det_{4s',\text{rep}}(\overline{\overline{{\mathbb M}}})
\det_{2[\perp{\bf\hat k}],\text{rep}}\left[{\sf Q}_{\bf\hat k}.
\left({\bf\hat k}.\overline{\overline{{\mathbb M}}}^{-1_{4s',
\text{rep}}}.{\bf\hat k}\right).{\sf Q}_{\bf\hat k}\right]
\right\}^{-1/2}.
\end{equation}

The determinants are once again straightforwardly obtained, 
and the limits $r\to 0$ and $\beta\to\infty$ carried out as 
in the previous section. Setting now 
\begin{mathletters}
\begin{eqnarray}
{\sf M}_1&=&{\sf Q}_{\bf\hat k}.({\bf\hat k}.{\mathbb M}_1^{-1_{4s'}}.
{\bf\hat k}).{\sf Q}_{\bf\hat k},\\
{\sf M}_2&=&{\sf Q}_{\bf\hat k}.({\bf k}.{\mathbb M}_1^{-1_{4s'}}\!:
\!{\mathbb M}_2\!:\!{\mathbb M}_1^{-1_{4s'}}.{\bf k}).{\sf Q}_{\bf\hat k},
\end{eqnarray} 
\end{mathletters}
the second-order perturbation expansion for the viscoplastic 
potential finally reads:
\begin{equation}
\label{psisigmasf}
\Psi_{\text{us}}(\Sigmasf)=\langle\psi_{\bf x}(\Sigmasf)\rangle
-{1\over 2}\int{d\Omega_{\bf\hat k}\over S_{d}}\left[
\mathop{\text{tr}}_{4}({\mathbb M}_2\!:\!{\mathbb M}_1^{-1_{4s'}})-
\mathop{\text{tr}}_{2}\left({\sf M}_2.
{\sf M}_1^{-1_{2[\perp{\bf\hat k}]}}\right)\right].
\end{equation}

%-----------------------------------------------------------------------------
\subsubsection{Application to the Norton law}
\label{attnlvp}
Writing $\omega_n=\langle \omega_n\rangle+\delta\omega_n$, and 
${\Sigmasfhat}={\Sigmasf}'/\sqrt{\mathop{\text{tr}}_2(\Sigmasf^{'2})}$, we have
\begin{mathletters}
\label{MM}
\begin{eqnarray}
{\mathbb M}_1&=&m_1 {\mathbb J}+m_2{\Sigmasfhat}
{\Sigmasfhat}=m_1 ({\mathbb J}-{\Sigmasfhat}
{\Sigmasfhat})+(m_1+m_2){\Sigmasfhat}
{\Sigmasfhat}\\
{\mathbb M}_2&=&m_3{\Sigmasfhat}{\Sigmasfhat},
\end{eqnarray}
\end{mathletters}
where
\begin{mathletters}
\label{mmm}
\begin{eqnarray}
m_1&=&{d\over d-1}\langle\omega_n\rangle \Sigma_{\text{eq}}^{n-1},\\
m_2&=&(n-1)m_1\\
m_3&=&{d\over d-1}\left\langle\delta\omega_n^2\right\rangle
\Sigma_{\text{eq}}^{2n}.
\end{eqnarray}
\end{mathletters}
Noting that ${\mathbb J}-{\Sigmasfhat}{\Sigmasfhat}$ 
and ${\Sigmasfhat}{\Sigmasfhat}$ are complementary 
orthogonal projectors in $L(4s')$, we compute ${\mathbb M}_1^{-1_{4s'}}$, 
${\mathbb M}_1^{-1_{4s'}}\!:\!{\mathbb M}_2$ and ${\mathbb M}_1^{-1_{4s'}}\!:
\!{\mathbb M}_2\!:\!{\mathbb M}_1^{-1_{4s'}}$. Now introducing the unit 
vector ${\bf\hat u}={\sf Q}_{\bf\hat k}.{\Sigmasfhat}.
{\bf\hat k}/|{\sf Q}_{\bf\hat k}.{\Sigmasfhat}.{\bf\hat k}|$, we 
express ${\sf M}_1$ and ${\sf M}_2$ in terms of ${\sf Q}_{\bf\hat k}-
{\bf\hat u}{\bf\hat u}$ and ${\bf\hat u}{\bf\hat u}$ and compute the 
rest. With
\begin{equation}
\label{sighatk}
\Sigma_{\bf\hat k}=1-2|{\sf Q}_{\bf\hat k}.{\Sigmasfhat}.
{\bf\hat k}|^2=1-2\left[{\bf\hat k}.{\Sigmasfhat}^2.{\bf\hat k}
-({\bf\hat k}.{\Sigmasfhat}.{\bf\hat k})^2\right],
\end{equation}
we arrive at
\begin{equation}
\label{simple}
\Psi_{\text{us}}(\Sigmasf)=\langle\psi_{\bf x}(\Sigmasf)\rangle
-{1\over 2}\int{d\Omega_{\bf\hat k}\over S_{d}}{m_3
\Sigma_{\bf\hat k}\over m_1+m_2\Sigma_{\bf\hat k}}.
\end{equation}
Finally replacing $m_1$, $m_2$, $m_3$ with their values (\ref{mmm}) 
yields the desired expression for the effective Norton viscoplastic 
potential:
\begin{mathletters}
\label{nortvis}
\begin{eqnarray}
\Psi_{\text{us}}(\Sigmasf)&=&{\omega_{n\,\text{eff}}(\Sigmasf)
\over n+1}\Sigma_{\text{eq}}^{n+1},\\
\omega_{n\,\text{eff}}(\Sigmasf)&=&\langle\omega_n\rangle
\left[1-{1\over 2}{\left\langle\delta\omega_n^2\right\rangle
\over \langle\omega_n\rangle^2}\int{d\Omega_{\bf\hat k}\over 
S_{d}}{(n+1)\Sigma_{\bf\hat k}\over 1+(n-1)\Sigma_{\bf\hat k}}
+O(\left\langle\delta\omega_n^3\right\rangle/\left\langle
\omega_n\right\rangle^3)\right].
\end{eqnarray}
\end{mathletters}

%-----------------------------------------------------------------------------
\section{Other types of disorder}
\label{otod}
We briefly discuss here how to cope with non site-disordered 
models, with possibly anisotropic correlations. It suffices 
to write the average of the exponential in (\ref{avzed}) 
simply as
\begin{equation}
\label{efpot}
\left\langle e^{-\beta \sum_{\gamma=1}^r\overline{\phi_{\bf x}
({\sf d}^\gamma)}}\right\rangle=\exp\left[\log\left\langle 
e^{-\beta \sum_{\gamma=1}^r\overline{\phi_{\bf x}({\sf d}^\gamma)}}\right
\rangle\right]
\end{equation}
and to expand the logarithm as in Sec.\ \ref{gfpd}, only keeping 
the quadratic terms in the outer exponential. Equ.\ 
(\ref{zutexpans}) is unchanged, save for a slightly different 
(but equivalent) form for the first averaged prefactor, and 
for a modification in the quadratic term in the main exponential. 
This term becomes \begin{equation}
-{1\over 2}{\beta\over V}\sum_{\alpha,\gamma}\int d{\bf x}\,d{\bf y}\, 
{\sf e}^\alpha({\bf x})\!:\! {\mathbb M}^{\alpha\gamma}({\bf x}|
{\bf y})\!:\! {\sf e}^\alpha({\bf y}),
\end{equation}
where
\begin{equation}
{\mathbb M}^{\alpha\gamma}({\bf x}|{\bf y})=\langle {\mathbb W}''
\rangle\delta_{\alpha\gamma}\delta({\bf x}-{\bf y})-{\beta\over V} 
\left[\langle{\sf W}'({\bf x}){\sf W}'({\bf y})\rangle-
\langle{\sf W}'({\bf x})\rangle\langle{\sf W}'({\bf y})\rangle\right].
\end{equation}
This requires modifications in the calculations, but does not 
change their principle. One merely has to deal with summations 
over another pair of indices, namely the coordinates ${\bf x}$ 
and ${\bf y}$. Now the determinants, e.g.\ in (\ref{zutnav}), 
become functional determinants and must simultaneously be 
evaluated as determinants of operators in the Fourier space, in 
addition to being operators on finite-dimensional vector spaces. 
However, this is not a serious problem in the perturbative 
calculation considered here since only traces are required 
in the final second-order results. 

In the case of translation-invariant disorder, the kernel 
${\mathbb M}^{\alpha\gamma}({\bf x}|{\bf y})$ is diagonal 
in the Fourier representation and (\ref{zutnav}) still holds, 
provided that one writes ${\mathbb M}({\bf k})$ (for its Fourier 
transform with respect to ${\bf x}-{\bf y}$) instead of 
${\mathbb M}$. The two-point correlation function $g({\bf x})$ 
is defined by:
\begin{mathletters}
\begin{eqnarray}
\left\langle \delta\theta_m({\bf x}) \delta\theta_m({\bf y})
\right\rangle&=&\left\langle \delta\theta_m^2\right\rangle 
g({\bf x}-{\bf y}),\\ 
\left\langle \delta\omega_n({\bf x}) \delta\omega_n({\bf y})
\right\rangle&=&\left\langle \delta\omega_n^2\right\rangle 
g({\bf x}-{\bf y}),
\end{eqnarray}
\end{mathletters}
with
\begin{equation}
\label{go}
g({\bf x}={\bf 0})=\int {d{\bf k}\over (2\pi)^d}\, g({\bf k})=1.
\end{equation}

The final results are then only modified as:
\begin{mathletters}
\begin{eqnarray}
\theta_{m\,\text{eff}}({\sf D})&=&\langle\theta_m\rangle
\left[1-{\left\langle\delta\theta_m^2\right\rangle\over 
\langle\theta_m\rangle^2}\int{d{\bf k}\over (2\pi)^d}\,
{(m+1) g({\bf k}) D_{\bf\hat k}\over 1+2(m-1)D_{\bf\hat k}}
+O(\left\langle\delta\theta_m^3\right\rangle/\left\langle
\theta_m\right\rangle^3)\right],\\
\omega_{n\,\text{eff}}(\Sigmasf)&=&\langle\omega_n\rangle
\left[1-{1\over 2}{\left\langle\delta\omega_n^2\right\rangle
\over \langle\omega_n\rangle^2}\int{d{\bf k}\over (2\pi)^d}
{(n+1) g({\bf k}) \Sigma_{\bf\hat k}\over 1+(n-1)
\Sigma_{\bf\hat k}}+O(\left\langle\delta\omega_n^3\right
\rangle/\left\langle\omega_n\right\rangle^3)\right].
\end{eqnarray}
\end{mathletters}
The Legendre duality property once again applies, 
whatever the nature of the correlations, and the result is once 
again independent of the boundary conditions. The previous 
expressions for isotropic site-disorder are recovered using 
$g({\bf x})=v\delta({\bf x})$. Ellipsoidal symmetry corresponds 
to a correlation function $g({\bf x})=g(|{\sf Z}.{\bf x}|)$, 
where ${\sf Z}$ is a symmetric constant tensor (Willis 1977; 
Ponte-Casta\~neda and Suquet, 1995).

%-----------------------------------------------------------------------------
\section{Discussion}
\label{disc}
In this section, we address the range of applicability of the 
functional method. Firstly, a major concern of studies on 
heterogeneous media is about bounds. In principle, the above 
functional formulation should be apt to deliver bounds, at 
least as far as the initial expressions (\ref{phiut}) and 
(\ref{psius}) are concerned. One might think, e.g.\ to use the 
convexity properties of the exponential in order to obtain 
inequalities on approximations to the potentials. However, 
to the knowledge of these authors, the replica method is an 
inadequate tool to use for subsequent calculations, if these 
are to preserve inequalities. Indeed, a convexity equality of 
the type
\begin{equation}
\mu\left(e^{\sum_{\alpha=1}^r {\cal V}^\alpha}\right)\ge 
e^{\mu\left(\sum_{\alpha=1}^r {\cal V}^\alpha \right)},
\end{equation}
where $\mu$ is some normalized functional measure and ${\cal V}$ 
is some potential, valid as long as the potential is replicated $r$ 
times, $r\ge 1$, may not survive the limit $r\to 0$. In the zero-replica 
limit indeed, quantities always positive when $r\geq 1$ may become
 negative, minima may transform into maxima (Parisi, 1984), and it 
is most often hard to conclude on inequalities. The replica method
 has been invented, and employed, to find estimates (which can be 
of very good quality) to free energies when disorder is present, not bounds.

Second, one must not confuse the initial functional formulation 
with the replica method, which we use as a tool in order to work 
out averages. These are distinct matters. Actually, there are 
two starting points for the calculations: either we transfer 
the statistical averages of the logarithm {\em inside} the 
functional integral by means of the replica method, and 
carry out approximations on the effective potential
\begin{equation}
-{1\over \beta}\log \left\langle e^{-\beta \sum_\gamma 
\overline{\phi_{\bf x}({\sf d}^\gamma)}}\right\rangle, 
\end{equation}
cf.\ (\ref{efpot}). This is what we did in the paper. 
Either we carry out an approximation on the non-replicated 
initial functional integral (\ref{phiut}), then take the 
logarithm, and average afterwards only. Save for 
perturbative expansions, the results will be different.

Third, all types of calculations feasible with classical methods 
(especially the perturbative ones) should translate into simple 
approximation schemes to the basic equations (\ref{phiut}) and 
(\ref{psius}). But we believe that the compact starting point 
of the functional formulation might enable one to obtain 
non-perturbative results more easily, and of different nature, 
than by classical means. There resides its main interest.

Finally, the trick consisting in writing the minimum of some 
functional as the limit of a particular functional integral 
could be applied elsewhere, e.g., to compute variational 
expressions such as those presented in Ref.\ \cite{CAST98} 
(computing a maximum can of course be done by identical means). 
 
%-----------------------------------------------------------------------------
\section{Conclusion}
\label{co}
A field-theoretic method in order to compute the effective 
properties of highly non-linear composites has been introduced. 
The minimization problem was shown to be equivalent to the 
computation of a functional integral. We believe the main 
advantage of this formulation is that it is valid for all 
types of local potentials, and that it can be used as a convenient
starting point for approximations. As a first application, a 
weak-disorder second-order perturbative calculation of this 
functional integral was presented. The second-order results of Suquet 
and Ponte-Casta\~neda (1993) for the effective plastic dissipation 
potential (with uniform traction boundary conditions) were recovered. 
In addition, the second-order correction to the viscoplastic potential 
was also obtained, and these results were extended to correlated 
disorder. With this method, minimizing over a vector velocity 
field, or a tensor stress field, is found to be equally feasible.

A natural extension of this work is to find a well-behaved non-perturbative 
approximation. Self-consistent calculations are currently under study.  

%-----------------------------------------------------------------------------
\acknowledgements
We thank P.\ Suquet and M.\ Garajeu for stimulating discussions, and for 
communicating to us very useful references, and J.-M.\ Diani for his 
interest in this work. Thanks are also due to Prof.\ J.\ R.\ Willis 
for useful comments.

%-----------------------------------------------------------------------------
\appendix

\section{Fourth-rank tensors in mechanics}
\label{ftim}
Before briefly reminding the reader about the main properties of fourth-rank 
tensors in mechanics, we need to set our notations about the various algebra 
and sub-algebras of linear operators that come into play. 

We designate by $L(2)$ the algebra of linear operators (second-rank 
$d\times d$ square matrices) ${\sf M}:{\bf R}^{d}\to {\bf R}^{d}$. 
Next, fourth-rank tensors appear in mechanics as representations of linear 
operators ${\mathbb M}:L(2)\to L(2)$. They can therefore be considered 
as square matrices, the indices of which are pairs of indices, and their 
components are denoted by $M_{ij,kl}$, such that $i$, $j$, $k$, $l=1,
\ldots,d$. We denote the algebra they belong to by $L(4)$. The product 
in $L(4)$ is
\begin{equation}
({\mathbb M}\!:\!{\mathbb N})_{ij,kl}=M_{ij,mn}N_{nm,kl}.
\end{equation}
Hence applying ${\mathbb M}$ to a second-rank tensor ${\sf N}$, we have 
\begin{equation}
({\mathbb M}\!:\!{\sf N})_{ij}=M_{ij,kl}N_{lk}.
\end{equation}
The identity ${\mathbb I}$ in $L(4)$ is 
\begin{equation}
I_{ij,kl}=\delta_{il}\delta_{jk},
\end{equation}
and the trace is defined (and denoted) by
\begin{equation}
{\text tr}_{4} {\mathbb M}=M_{ij,ji}.
\end{equation}
Hence ${\text tr}_{4}{\mathbb I}=d^2$.

The algebra $L(2)$ can be split in the direct sum of that of 
symmetric traceless matrices [which we denote by $L(2s')$], 
diagonal matrices proportional to the identity [$L(2d)$], and anti-symmetric 
matrices [$L(2a)$]. The algebra of symmetric matrices is the direct sum 
$L(2s)=L(2s')\oplus L(2d)$. The algebra $L(4)$ therefore admits three 
remarkable sub-algebras, which we denote by $L(4s')$, $L(4d)$, $L(4a)$, 
generated by the mutually orthogonal projectors ${\mathbb J}$, 
${\mathbb K}$ and ${{\mathbb I}}^a$ respectively, defined by
\begin{mathletters}
\begin{eqnarray}
J_{ij,kl}&=&{1\over 2}(\delta_{il}\delta_{jk}+\delta_{ik}\delta_{jl})
-{1\over d}\delta_{ij}\delta_{kl},\\
K_{ij,kl}&=&{1\over d}\delta_{ij}\delta_{kl},\\
I^a_{ij,kl}&=&{1\over 2}(\delta_{il}\delta_{jk}-\delta_{ik}\delta_{jl}).
\end{eqnarray}
\end{mathletters}
They obey ${\mathbb I}={\mathbb I}^a+{\mathbb J}+{\mathbb K}$. Hence, 
the sub-algebras $L(4s')$, $L(4d)$, $L(4a)$ are that of the 
endomorphisms on $L(2s')$, $L(2d)$, $L(2a)$ respectively. The 
operators ${\mathbb J}$, ${\mathbb K}$, and ${\mathbb I}^a$ are 
the appropriate identity operators in each sub-algebra. We have
\begin{mathletters}
\begin{eqnarray}
{\text tr}_{4}{\mathbb J}&=&d(d+1)/2-1,\\
{\text tr}_{4}{\mathbb K}&=&1,\\
{\text tr}_{4}{\mathbb I}^a&=&d(d-1)/2.
\end{eqnarray}
\end{mathletters}
These numbers count the number of eigenvalues of each operator. 
In $L(4s)$, the sub-algebra of the endomorphisms on $L(2s)$, the 
identity is ${\mathbb I}^s={\mathbb J}+{\mathbb K}$:
\begin{equation}
I^s_{ij,kl}={1\over 2}(\delta_{il}\delta_{jk}+\delta_{ik}\delta_{jl}),
\end{equation}
and ${\text tr}_{4}{\mathbb I}^s=d(d+1)/2$. Note that $L(4s')
\oplus L(4d)\subset L(4s)$, but the two sets are not equal.

Finally note that the inverse (resp.\ determinant) in $L(4)$ of 
an operator ${\mathbb M}\in L(4s')\oplus L(4d)\oplus L(4a)$ is 
the sum (resp.\ product) of the inverses (reps.\ determinants) of 
its constituents in their respective sub-algebra.

%-----------------------------------------------------------------------------
\section{Gaussian integrals}
\label{gi}
\subsection{Gaussian integrals over vector fields}
\label{giovf}
We start from the well-known $d$-dimensional gaussian integral 
over a vector field ${\lambdabf}$ (Kleinert, 1995). If ${\sf M}$ 
is a $d\times d$ positive-definite symmetric matrix, and ${\bf b}$ 
any vector, then
\begin{equation}
\label{gaussvec}
\int d{\lambdabf} \,\exp\left(-{1\over 2}\lambdabf.{\sf M}.\lambdabf
+{\bf b}.\lambdabf \right)=\left[{(2\pi)^d\over\det_2({\sf M})}\right]^{1/2}
\exp\left({1\over 2}{\bf b}.{\sf M}^{-1_2}.{\bf b}\right),
\end{equation}
where the integration measure is $d{\lambdabf}=\prod_i d\lambda_i$.

This formula is meaningless if ${\sf M}$ is not invertible in $L(2)$. 
However, let ${\bf\hat k}$ be a unit vector, ${\bf b}$ now be a real 
vector, and ${\sf M}={\sf M}_\perp+\epsilon^2 {\bf\hat k}{\bf\hat k}$, 
where ${\sf M}_\perp$ is invertible only in $L(2{[\perp{\bf\hat k}]})$, 
the set of the endomorphisms which operate in the subspace of vectors 
{\em orthogonal} to ${\bf\hat k}$. Then ${\sf M}^{-1_2}=
{\sf M}_\perp^{-1_{\smash{2{[\perp{\bf\hat k}]}}}}+\epsilon^{-2} 
{\bf\hat k}{\bf\hat k}$, $\det_2({\sf M})=\epsilon^{2}
\det_{2{[\perp{\bf\hat k}]}}({\sf M}_\perp)$, and
\begin{equation}
\int d{\lambdabf} \,\exp\left(-{1\over 2}\lambdabf.{\sf M}.
\lambdabf+i{\bf b}.\lambdabf \right)=\left[{(2\pi)^d\over
\epsilon^{2}\det_{2{[\perp{\bf\hat k}]}}({\sf M}_\perp)}\right]^{1/2}
\exp\left[-{1\over 2}{\bf b}.
{\sf M}^{-1_{\smash{2{[\perp{\bf\hat k}]}}}}.{\bf b}
-{1\over 2\epsilon^2}({\bf b}.{\bf\hat k})^2\right].
\end{equation}
Letting $\epsilon\to 0$ and using the representation of the 
Dirac distribution
\begin{equation}
\label{dirac}
\delta(x)=\lim_{\epsilon\to 0}{1\over\sqrt{2\pi\epsilon^2}}
e^{-{1\over 2}(x/\epsilon)^2},
\end{equation}
we obtain
\begin{equation}
\label{ilam}
\int d{\lambdabf} \,\exp\left(-{1\over 2}\lambdabf.
{\sf M}_\perp.\lambdabf+i{\bf b}.\lambdabf \right)=
\left[{(2\pi)^{d+1}\over\det_{2{[\perp{\bf\hat k}]}}
({\sf M}_\perp)}\right]^{1/2}\exp\left(-{1\over 2}{\bf b}.
{\sf M}^{-1_{2{[\perp{\bf\hat k}]}}}.{\bf b}\right)
\delta({\bf b}.{\bf\hat k}).
\end{equation}
We define the integration measure over {\em the vectors 
orthogonal to} ${\bf\hat k}$ to be
\begin{equation}
\label{vecperp}
\mathop{d{\lambdabf}}_{[\perp{\bf\hat k}]}=\delta(\lambdabf.
{\bf\hat k})d{\lambdabf}.
\end{equation}
Multiplying both sides of (\ref{ilam}) by $\exp(i{\bf b}.
{\bf a})$, where ${\bf a}$ is a real vector, and integrating 
over ${\bf b}$ yields
\begin{eqnarray}
&&\int \mathop{d{\bf b}}_{[\perp{\bf\hat k}]}\exp
\left(-{1\over 2}{\bf b}.{\sf M}^{-1_{\smash{2{[\perp{\bf\hat k}]}}}}.
{\bf b}+i{\bf b}.{\bf a}\right)\nonumber\\
&=&\left[{(2\pi)^{-(d+1)}\over\det_{2{[\perp{\bf\hat k}]}}
({\sf M}_\perp^{-1_{\smash{2{[\perp{\bf\hat k}]}}}})}\right]^{1/2}
\int d{\lambdabf}d{\bf b} \,\exp\left[-{1\over 2}\lambdabf.
{\sf M}_\perp.\lambdabf+i{\bf b}.(\lambdabf +{\bf a})\right]
\nonumber\\
&=&(2\pi)^d\left[{(2\pi)^{-(d+1)}\over\det_{2{[\perp{\bf\hat k}]}}
({\sf M}_\perp^{-1_{\smash{2{[\perp{\bf\hat k}]}}}})}\right]^{1/2}
\int d{\lambdabf}\,\exp\left(-{1\over 2}\lambdabf.{\sf M}_\perp.
\lambdabf\right)\delta({\bf a}+\lambdabf)\nonumber\\
\label{ilam2}
&=&\left[{(2\pi)^{d-1}\over\det_{2{[\perp{\bf\hat k}]}}
({\sf M}_\perp^{-1_{\smash{2{[\perp{\bf\hat k}]}}}})}\right]^{1/2}
\exp\left(-{1\over 2}{\sf a}.{\sf M}_\perp.{\sf a}\right),
\end{eqnarray}
where $\delta({\bf x})=\prod_i\delta(x_i)$. Note that we can 
exchange ${\sf M}_\perp^{-1_{2{[\perp{\bf\hat k}]}}}$ and ${\sf M}_\perp$ 
in (\ref{ilam2}). Moreover, ${\sf M}_\perp$ can be written 
${\sf M}_\perp={\sf Q}_{\bf\hat k}.{\sf M}.{\sf Q}_{\bf\hat k}$, 
where ${\sf Q}_{\bf\hat k}$ is the projector ${\sf Q}_{\bf\hat k}
={\sf I}-{\bf\hat k}{\bf\hat k}$, for some ${\sf M}\in L(2)$. 
Besides, the ${\bf b}$ are now orthogonal to ${\bf\hat k}$ (because of 
the integration measure) so that ${\bf b}.{\sf Q}_{\bf\hat k}.{\sf M}.
{\sf Q}_{\bf\hat k}.{\bf b}={\bf b}.{\sf M}.{\bf b}$. Finally, the 
result can be analytically continued to complex values of ${\bf a}$. 
Changing the names of the variables, we therefore arrive at the formula
\begin{equation}
\label{lamperp}
\int \mathop{d{\lambdabf}}_{[\perp{\bf\hat k}]} \,\exp\left(-{1\over 2}
\lambdabf.{\sf M}.\lambdabf+{\bf b}.\lambdabf \right)=\left[{(2\pi)^{d-1}
\over\det_{2{[\perp{\bf\hat k}]}}({\sf Q}_{\bf\hat k}.
{\sf M}.{\sf Q}_{\bf\hat k})}\right]^{1/2}\exp\left[{1\over 2}
{\bf b}.({\sf Q}_{\bf\hat k}.{\sf M}.
{\sf Q}_{\bf\hat k})^{-1_{2{[\perp{\bf\hat k}]}}}.{\bf b}\right].
\end{equation}

%-----------------------------------------------------------------------------
\subsection{Gaussian integrals over tensors of rank 2}
\label{giotor2}
The generic gaussian integral over all matrices ${\sf s}\in L(2)$ is
\begin{equation}
\label{gaussmat}
\int d{\sf s}\, \exp\left(-{1\over2} {\sf s}\!:\!{\mathbb M}\!:
\!{\sf s}+{\sf b}\!:\!{\sf s}\right)=\left[{(2\pi)^{d^2}\over 
\det_4({\mathbb M})}\right]^{1/2}\exp\left({1\over2} 
{\sf b}\!:\!{\mathbb M}^{-1_4}\!:\!{\sf b}\right),
\end{equation}
where $d{\sf s}=\prod_{ij}ds_{ij}$ is the appropriate measure, 
${\mathbb M}$ is a symmetric matrix of $L(4)$, i.e.\ is such 
that $M_{ij,kl}=M_{kl,ij}$. This formula is a direct consequence 
of (\ref{gaussvec}) via a mapping of ${\sf s}$ onto a column 
vector in $\smash{{\bf R}^{d^2}}$. Once again, the integral 
is meaningless in general if ${\mathbb M}$ is not invertible 
in $L(4)$. We apply the same method as in Sec.\ \ref{giovf}.

%-----------------------------------------------------------------------------
\subsubsection{Gaussian integrals over symmetric matrices}
\label{giosm}
Let ${\sf b}\in L(2)$ be real. We consider ${\mathbb M}={\mathbb M}_s+
(\epsilon^2/2) {\mathbb I}^a$, where ${\mathbb M}_s\in L(4s)$ and is 
invertible in $L(4s)$. Then ${\mathbb M}^{-1_4}={\mathbb M}_s^{-1_{4s}}+
(2/\epsilon^2){\mathbb I}^a$ and $\det_4({\mathbb M})=\det_{4s}
({\mathbb M})(\epsilon^2/2)^{d(d-1)/2}$. Hence from (\ref{gaussmat}), 
\begin{eqnarray}
&&\qquad\int d{\sf s}\,\exp\left(-{1\over2} {\sf s}\!:
\!{\mathbb M}\!:\!{\sf s}+i{\sf b}\!:\!{\sf s}\right)\nonumber\\
&&{}=\left[{(2\pi)^{d^2}\over \det_{4s}({\mathbb M}_s)}\right]^{1/2}
{1\over (\epsilon^2/2)^{d(d-1)/4}}\exp\left(-{1\over2} {\sf b}\!:
\!{\mathbb M}_s^{-1_{4s}}\!:\!{\sf b}-{1\over\epsilon^2} {\sf b}\!:
\!{\mathbb I}^a\!:\!{\sf b}\right).
\end{eqnarray}
But ${\sf b}\!:\!{\mathbb I}^a\!:\!{\sf b}=(1/2)\sum_{i<j}(b_{ij}-b_{ji})^2$. 
Using (\ref{dirac}) and going to the limit $\epsilon\to 0$, we find
\begin{eqnarray}
\label{notinv1}
&&\qquad\int d{\sf s}\, \exp\left(-{1\over2} {\sf s}\!:
\!{\mathbb M}_s\!:\!{\sf s}+i{\sf b}\!:
\!{\sf s}\right)\nonumber\\
&&{}=\left[{(2\pi)^{d^2}(4\pi)^{d(d-1)/2}\over 
\det_{4s}({\mathbb M}_s)}\right]^{1/2}\exp\left(-{1\over2} {\sf b}\!:
\!{\mathbb M}_s^{-1_{4s}}\!:\!{\sf b}\right)\prod_{i<j}\delta(b_{ij}-b_{ji}).
\end{eqnarray}

We define the measure over the {\em symmetric} tensors of rank two as:
\begin{equation}
\label{symmes}
d_s{\sf s}=2^{d(d-1)/4}[\prod_{i,j<i} \delta(s_{ij}-s_{ji})]\,d{\sf s}.
\end{equation}
Multiplying both sides of (\ref{notinv1}) by $\exp(i {\sf b}\!:\!
{\sf a})$, and integrating with respect to ${\sf b}$ with 
$d{\sf b}$ yields
\begin{eqnarray}
&&\int d_s{\sf b}\, \exp\left(-{1\over2} {\sf b}\!:
\!{\mathbb M}_s^{-1_{4s}}\!:\!{\sf b}+i{\sf a}\!:\!{\sf b}\right)\nonumber\\
&=&\left[{(2\pi)^{-d^2-d(d-1)/2}\over 
\det_{4s}({\mathbb M}_s^{-1_{4s}})}\right]^{1/2}\int 
d{\sf b}\,d{\sf s}\,\exp\left(-{1\over2} {\sf s}\!:\!{\mathbb M}_s\!:
\!{\sf s}+i({\sf s}+{\sf a})\!:\!{\sf b}\right)\nonumber\\
&=&(2\pi)^{d^2}\left[{(2\pi)^{-d^2-d(d-1)/2}\over \det_{4s}
({\mathbb M}_s^{-1_{4s}})}\right]^{1/2}\int d{\sf s}\,\exp
\left(-{1\over2} {\sf s}\!:\!{\mathbb M}_s\!:\!{\sf s}\right)
\delta({\sf s}+{\sf a})\nonumber\\
&=&\left[{(2\pi)^{d(d+1)/2}\over \det_{4s}({\mathbb M}_s^{-1_{4s}})}
\right]^{1/2}\exp\left(-{1\over2} {\sf a}\!:\!{\mathbb M}_s\!:
\!{\sf a}\right),
\end{eqnarray}
where $\delta({\sf x})=\prod_{ij}\delta(x_{ij})$. Whence the 
generic result for gaussian integrals over {\em symmetric} 
matrices, with ${\mathbb M}\in L(4)$:
\begin{equation}
\label{gaussym}
\int d_s{\sf s}\, \exp\left(-{1\over2} {\sf s}\!:\!{\mathbb M}\!:
\!{\sf s}+{\sf b}\!:\!{\sf s}\right)=\left[{(2\pi)^{d(d+1)/2}\over 
\det_{4s}({\mathbb I}^s\!:\!{\mathbb M}\!:
\!{\mathbb I}^s)}\right]^{1/2}\exp\left({1\over2} {\sf b}\!:
\!({\mathbb I}^s\!:\!{\mathbb M}\!:\!{\mathbb I}^s)^{-1_{4s}}\!:
\!{\sf b}\right).
\end{equation}

%-----------------------------------------------------------------------------
\subsubsection{Gaussian integrals over traceless symmetric matrices}
\label{giotsm}
If ${\mathbb I}^s\!:\!{\mathbb M}\!:\!{\mathbb I}^s$ is not 
invertible in $L(4s)$, but only in $L(4s')$ for instance, the 
procedure can be repeated: let us assume that ${\mathbb M}=
{\mathbb M}_{s'}+\epsilon^2{\mathbb K}$, where ${\mathbb M}_{s'}
\in L(4s')$. Starting from (\ref{gaussym}) with ${\sf b}\to i{\sf b}$, 
using $({\mathbb I}^s\!:\!{\mathbb M}\!:\!{\mathbb I}^s)^{-1_{4s}}
={\mathbb M}_{s'}^{-1_{4s'}}+{\mathbb K}/\epsilon^2$, $\det_{4s}
({\mathbb I}^s\!:\!{\mathbb M}\!:\!{\mathbb I}^s)=\det_{4s'}
({\mathbb M}_{s'})\epsilon^2$ and letting $\epsilon\to 0$, one 
finds:
\begin{equation}
\label{notinv2}
\int d_s{\sf s}\, \exp\left(-{1\over2} {\sf s}\!:\!{\mathbb M}_{s'}\!:
\!{\sf s}+i{\sf b}\!:\!{\sf s}\right)=\left[{d(2\pi)^{d(d+1)/2+1}\over 
\det_{4s'}({\mathbb M}_{s'})}\right]^{1/2}\exp\left(-{1\over2} {\sf b}\!:
\!{\mathbb M}_{s'}^{-1_{4s'}}\!:\!{\sf b}\right)\delta(tr_2{\sf b}).
\end{equation}

We define the integration measure over {\em symmetric traceless} 
tensors to be
\begin{equation}
 d_{s'}{\sf s}=d^{1/2}\delta(tr_2{\sf s})\,d_s{\sf s}.
\end{equation}
As one easily checks, we have, with (\ref{symmes})
\begin{equation}
\int d_s{\sf x}\, \exp(i {\sf s}\!:\!{\sf x})= 2^{d(d-1)/4}(2\pi)^{d(d+1)/2}
[\prod_i\delta(s_{\underline{i}\underline{i}})][\prod_{i<j}\delta(s_{ij}+s_{ji})].
\end{equation}
Multiplying (\ref{notinv2}) by $\exp(i{\sf a}\!:\!{\sf b})$, and 
integrating over ${\sf b}$ with measure $d_s{\sf b}$, we deduce, 
using ${\mathbb M}_{s'}={\mathbb J}\!:\!{\mathbb M}\!:\!{\mathbb J}$, 
the result for gaussian integrals over {\em symmetric traceless} 
tensors: 
\begin{equation}
\label{gaussymts}
\int d_{s'}{\sf s}\, \exp\left(-{1\over2} {\sf s}\!:\!{\mathbb M}\!:
\!{\sf s}+{\sf b}\!:\!{\sf s}\right)=\left[{(2\pi)^{d(d+1)/2-1}\over 
\det_{4s'}({\mathbb J}\!:\!{\mathbb M}\!:\!{\mathbb J})}\right]^{1/2}
\exp\left[{1\over2} {\sf b}\!:\!({\mathbb J}\!:\!{\mathbb M}\!:
\!{\mathbb J})^{-1_{4s'}}\!:\!{\sf b}\right].
\end{equation}

%-----------------------------------------------------------------------------
\subsubsection{Other gaussian integrals}
\label{ogi}
For the sake of completeness, we finally give without demonstration 
the results for gaussian integrals over {\em antisymmetric tensors}, 
with measure
\begin{equation}
d_a{\sf s}=2^{d(d-1)/4}[\prod_{i}\delta(s_{\underline{i}\underline{i}})][\prod_{i<j}
\delta(s_{ij}+s_{ji})]d{\sf s},
\end{equation}
and over {\em diagonal tensors proportional to the identity}, of 
the type $s{\sf I}/d$, with measure
\begin{equation}
d_d{\sf s}=d^{-1/2}d{\sf s}\int ds\,\delta({\sf s}-s{\sf I}/d).
\end{equation}
One finds
\begin{eqnarray}
\label{gaussasym}
\int d_a{\sf s}\, \exp\left(-{1\over2} {\sf s}\!:\!{\mathbb M}\!:
\!{\sf s}+{\sf b}\!:\!{\sf s}\right)&=&\left[{(2\pi)^{d(d-1)/2}\over 
\det_{4a}({\mathbb I}^a\!:\!{\mathbb M}\!:\!{\mathbb I}^a)}\right]^{1/2}
\exp\left[{1\over2} {\sf b}\!:\!({\mathbb I}^a\!:\!{\mathbb M}\!:
\!{\mathbb I}^a)^{-1_{4a}}\!:\!{\sf b}\right]\\
\label{gaussdiag}
\int d_d{\sf s}\, \exp\left(-{1\over2} {\sf s}\!:\!{\mathbb M}\!:
\!{\sf s}+{\sf b}\!:\!{\sf s}\right)&=&
\left[{2\pi\over \det_{4d}({\mathbb K}\!:\!{\mathbb M}
\!:\!{\mathbb K})}\right]^{1/2}\exp\left[{1\over2} {\sf b}\!:
\!({\mathbb K}\!:\!{\mathbb M}\!:\!{\mathbb K})^{-1_{4d}}\!:\!{\sf b}\right].
\end{eqnarray}
The last identity is trivial (scalar gaussian integral).

%-----------------------------------------------------------------------------
\section{Inverse and determinant in replica space}
\label{iadirs}
In replica space, $A$ and $B$ being replica-independent operators 
pertaining to some algebra ${\cal A}$, the inverse and determinant 
of a matrix of the type
\begin{equation}
\overline{\overline{M}}=A\overline{\overline{I}}+B\overline{\overline{U}},
\end{equation}
read
\begin{mathletters}
\label{invdet}
\begin{eqnarray}
\label{inv}
\overline{\overline{M}}^{-1_{{\cal A},\text{rep}}}&=&
A^{-1_{\cal A}}\overline{\overline{I}}-(A+rB)^{-1_{\cal A}}
BA^{-1_{\cal A}}\overline{\overline{U}},\\
\label{det}
\det_{{\cal A},\text{rep}}(\overline{\overline{M}})&=&
\det_{\cal A}(A)^{r-1}\det_{\cal A}(A+rB)=1+r\left[\log\det_{\cal A}
(A)+\mathop{\text{tr}}_{\cal A}(BA^{-1_{\cal A}})\right]+O(r^2)
\end{eqnarray}
\end{mathletters}
The expansion derives from the operator identity (Kleinert, 1995)  
$\log\det=\mathop{\text{tr}}\log$.

%-----------------------------------------------------------------------------
\section{Gaussian integrals over the velocity and stress fields}
\label{giotvasf}
\subsection{Integration over the velocity field}
\label{iotvf}
We detail here the steps leading to (\ref{zutnav}).

The Dirac distribution implementing the constraint 
$\nablabf.{\bf v}=0$ in (\ref{sms}) is exponentiated 
first: we introduce a scalar field $\lambda$ and write:
\begin{equation}
\delta(\nablabf.{\bf v})\propto\int {\cal D}
\lambda\,\exp\left[i\int\! d{\bf x}\,\lambda({\bf x})
\nablabf.{\bf v}({\bf x})\right].
\end{equation}
Then we Fourier transform the integrand. By definition, for any 
function $f$ (which will be ${\mathbf\lambda}$ or ${\sf s}$), 
the Fourier transform is:
\begin{equation}
f({\bf k})=\int\!d{\bf x}\,e^{i{\bf k}.{\bf x}} f({\bf x}),
\quad f({\bf x})=\int\!\frac{d{\bf k}}{(2\pi)^d}\,e^{-i{\bf k}.
{\bf x}} f({\bf k}).
\end{equation}
Since $f({\bf x})$ is real, we write for ${\bf k}\not={\bf 0}$:
\begin{mathletters}
\begin{eqnarray}
f({\bf k})&=&{1\over \sqrt{2}}[f^{(1)}({\bf k})+if^{(2)}({\bf k})],\\
f(-{\bf k})&=&{1\over \sqrt{2}}[f^{(1)}({\bf k})-if^{(2)}({\bf k})],
\end{eqnarray}
\end{mathletters}
where $f^{(1)}$ and $f^{(2)}$ are real functions of the 
wavevector; also, we set $f({\bf k}={\bf 0})=f^{(1)}({\bf 0})$. 
This decomposition is carried out for all $\lambda^\alpha({\bf k})$ 
and ${\bf v}^\alpha({\bf k})$.

Introducing the set ${\cal K}^+=\{{\bf k}/k_1>0\}\
\cup\{{\bf k}/k_1=0,k_2>0\}\cup\ldots\cup\{{\bf k}/k_1=0,
\ldots,k_{d-1}=0,k_d>0\}$, a functional measure ${\cal D}f$ 
over $f({\bf x})$ thus becomes in the Fourier representation
\begin{equation}
{\cal D}f=\prod_{\bf x}df({\bf x})\propto df^{(1)}({\bf 0})
\prod_{{\bf k}\in{\cal K}^+} [df^{(1)}({\bf k})df^{(2)}({\bf k})].
\end{equation}
Note that $f^{(1)}$ and $f^{(2)}$ are defined only for 
${\bf k}\in{\cal K}^+$. By Parseval's identity, we have
\begin{equation}
\int\! d{\bf x}\,\lambda({\bf x})\nablabf.{\bf v}({\bf x})
=\int_{{\cal K}^+}\!\frac{d{\bf k}}{(2\pi)^d}\left[\,\lambda^{(2)}
({\bf k}){\bf v}^{(1)}({\bf k}).{\bf k}-\lambda^{(1)}({\bf k})
{\bf v}^{(2)}({\bf k}).{\bf k}\,\right].
\end{equation}
As usual in functional methods, the latter integral is to be 
understood as a Riemann discrete sum by applying 
the correspondence:
\begin{equation}
\int \frac{d{\bf k}}{(2\pi)^d}\to \frac{1}{V}\sum_{\bf k}.
\end{equation}
In particular, this allows one to separate the contribution of 
the mode ${\bf k}=0$. The argument of the exponential in (\ref{sms}) 
is likewise transformed into a sum of Fourier modes via
\begin{eqnarray}
&&\qquad\int\!d{\bf x}\,{\sf e}^\alpha({\bf x})\!:
\!{\mathbb M}^{\alpha\gamma}\!:\!{\sf e}^\gamma({\bf x})\nonumber\\
&=&{1\over V}\,{\sf e}^{\alpha(1)}({\bf 0})\!:
\!{\mathbb M}^{\alpha\gamma}\!:\!{\sf e}^{\gamma(1)}({\bf 0})
+{1\over V}\sum_{{\bf k}\in{\cal K}^+}\left[{\sf e}^{\alpha(1)}
({\bf k})\!:\!{\mathbb M}^{\alpha\gamma}\!:\!{\sf e}^{\gamma(1)}
({\bf k})+{\sf e}^{\alpha(2)}({\bf k})\!:\!{\mathbb M}^{\alpha\gamma}
\!:\!{\sf e}^{\gamma(2)}({\bf k})\right]\nonumber\\
&=&{1\over V}\sum_{{\bf k}\in{\cal K}^+}\left[{\bf v}^{\alpha(1)}
({\bf k}).({\bf k}.{\mathbb M}^{\alpha\gamma}.{\bf k}).
{\bf v}^{\gamma(1)}({\bf k})+{\bf v}^{\alpha(2)}({\bf k}).
({\bf k}.{\mathbb M}^{\alpha\gamma}.{\bf k}).
{\bf v}^{\gamma(2)}({\bf k})\right].
\end{eqnarray}
In the last equality, we have used the symmetry of 
${\mathbb M}^{\alpha\gamma}$ with respect to its tensor indices. 
Note that ${\sf e}^{(1)}=-[{\bf k}{\bf v}^{(2)}+{\bf v}^{(2)}
{\bf k}]/2$ and ${\sf e}^{(2)}=[{\bf k}{\bf v}^{(1)}+
{\bf v}^{(1)}{\bf k}]/2$.

Because of the constraints $\delta(\overline{\sf e}^\alpha)$, the 
linear terms $\overline{\sf e}^\alpha$ disappear from the 
exponential. There is no dependence of the integrand with respect 
to $\lambda^{(1)}({\bf 0})$ nor to ${\bf v}^{(1)}({\bf 0})$, so 
that integration over these variables only yields harmless 
$\beta$-independent infinite multiplicative factors which we 
drop out, since they do not contribute to the final result 
in the limit $\beta\to\infty$. The reason is the same as that 
invoked in conjunction with Equ.\ (\ref{simplif}). Such 
physically irrelevant (because multiplicative) infinities 
are often encountered when dealing with functional integrals. 
They are related to (here) unimportant normalization questions.

We therefore see that for each ${\bf k}\not={\bf 0}$, mutually 
complex conjugate partial integrals $A({\bf k})$ $=$ $\int$ 
$d\lambda^{\alpha(2)}({\bf k})$ $d{\bf v}^{\alpha(1)}({\bf k})$ 
$[\ldots]$ and $A^*({\bf k})=\int d\lambda^{\alpha(1)}({\bf k}) 
d{\bf v}^{\alpha(2)}({\bf k})[\ldots]$ show up in pairs, and 
can be evaluated independently from one another. Renaming, 
in $A({\bf k})$, the integration variables into 
${\bf k}$-independent $d\lambda^\alpha$ and 
$d{\bf v}^\alpha$, we arrive at
\begin{equation}
\left\langle Z_{\text{ut}}^r\right\rangle\propto
\langle e^{-r\beta {v\over V} \phi_{\bf x}({\sf D})}
\rangle^{V\over v}\prod_{{\bf k}\in{\cal K}^+}
 |A({\bf k})|^2,
\end{equation}
where
\begin{mathletters}
\begin{eqnarray}
\label{ligne11}
A({\bf k})&=&\int\left(\prod_{\alpha} 
d\lambda^\alpha d{\bf v}^\alpha\right)
\exp\left[-{1\over 2}{\beta\over V^2} \sum_{\alpha\gamma}
 {\bf v}^\alpha .({\bf k}.{\mathbb M}^{\alpha\gamma}.{\bf k}).
{\bf v}^\gamma+{i\over V}\sum_\alpha 
\lambda^\alpha ({\bf v}^\alpha.{\bf k})\right]\\
\label{ligne21}
&\propto&\int\left[\prod_{\alpha} d{\bf v}^\alpha 
\delta({\bf v}^\alpha.{\bf k})\right]
\exp\left[-{1\over 2}{\beta\over V^2} 
\sum_{\alpha\gamma} {\bf v}^\alpha .({\bf k}.
{\mathbb M}^{\alpha\gamma}.{\bf k}).{\bf v}^\gamma\right]\\
\label{ligne31}
&\propto&\det_{2[\perp{\bf\hat k}],\text{rep}}
\left[{\sf Q}_{\bf\hat k}.\left({\bf\hat k}.
\overline{\overline{{\mathbb M}}}.{\bf\hat k}\right).
{\sf Q}_{\bf\hat k}\right]^{-1/2},
\end{eqnarray}
\end{mathletters}
and ${\sf Q}_{\bf\hat k}={\sf I}-{\bf\hat k}{\bf\hat k}$ 
with ${\bf\hat k}={\bf k}/k$. In the step from 
(\ref{ligne21}) to (\ref{ligne31}) we used 
formula (\ref{lamperp}), trivially generalized 
to the replica space, and discarded, among other 
irrelevant factors, a power of the modulus $k$. 
Equ.\ (\ref{zutnav}) finally follows from the 
fact that $A({\bf k})=A(-{\bf k})$.

%-----------------------------------------------------------------------------
\subsection{Integration over the stress field}
\label{iotsf}
We now detail the steps leading to (\ref{znav}), which 
differ little from above. Because of our considering 
the Norton viscoplastic potential (which does not 
depend on ${\bf \sigma}_{\text{m}}$) for explicit 
applications, the matrix ${\mathbb M}_1$ given in 
(\ref{MM}) is not invertible in $L(4s)$, but only in 
$L(4s')$. As a consequence [cf.\ (\ref{inv})], 
$\overline{\overline{\mathbb M}}$ is not invertible 
in $L(4s,\text{rep})$, but only in $L(4s',\text{rep})$. 
This leads to peculiarities that have to be taken into 
account but does not change the principle of the 
calculation. The equivalent of formula (\ref{znav}) 
for $\overline{\overline{\mathbb M}}$ invertible 
in $L(4s,\text{rep})$ is provided hereafter for 
the sake of completeness. 

The Dirac distribution implementing the constraint 
$\nablabf.{\sf s}={\bf 0}$ in the measure 
$\tilde{\cal D}_{s}{\sf s}$ is exponentiated by 
the introduction of a vector field ${\lambdabf}$ as:
\begin{eqnarray}
\delta(\nablabf.{\sf s})&\propto&\int {\cal D}{\lambdabf}
\,\exp\left(i\int\! d{\bf x}\,{\lambdabf}({\bf x}).
[\nablabf.{\sf s}({\bf x})]\right)\nonumber\\
&=&\int {\cal D}{\lambdabf}\,\exp\left(i{1\over V}
\sum_{{\bf k}\in{\cal K}+}\left\{{\sf s}^{(1)}({\bf k})
\!:\![{\bf k}\,{\lambdabf}^{(2)}({\bf k})]-{\sf s}^{(2)}
({\bf k})\!:\![{\bf k}\,{\lambdabf}^{(1)}({\bf k})]\right\}\right).
\end{eqnarray}
A reasoning paralleling that in Sec.\ \ref{iotvf} yields:
\begin{equation}
\left\langle Z_{\text{us}}^r\right\rangle\propto
\langle e^{-r\beta {v\over V} \psi_{\bf x}(\Sigmasf)}
\rangle^{V\over v}\prod_{{\bf k}\in{\cal K}^+} 
|A({\bf k})|^2,
\end{equation}
where now
\begin{mathletters}
\begin{eqnarray}
\label{ligne1}
A({\bf k})&=&\int\left(\prod_{\alpha} 
d\lambdabf^\alpha d_s{\sf s}^\alpha\right)
\exp\left[-{1\over 2}{\beta\over V^2} 
\sum_{\alpha\gamma} {\sf s}^\alpha \!:\!{\mathbb M}^{\alpha\gamma}
\!:\!{\sf s}^\gamma+{i\over V}\sum_\alpha {\sf s}^\alpha
\!:\!({\bf k}\,\lambdabf^\alpha)\right]\\
\label{ligne2}
&\propto&\det_{4s',\text{rep}}
(\overline{\overline{{\mathbb M}}})^{-1/2}\int\left(\prod_{\alpha}
\mathop{d\lambdabf^\alpha}_{[\perp{\bf\hat k}]}\right)
\exp\left\{-{1\over 2}{k^2\over\beta}\sum_{\alpha\gamma} 
\lambdabf^\alpha.\left[{\bf\hat k}.
\left({\mathbb M}^{-1_{4s',\text{rep}}}\right)^{\alpha\gamma}.
{\bf\hat k}\right].\lambdabf^\gamma\right\}\\
\label{ligne3}
&\propto&\det_{4s',\text{rep}}
(\overline{\overline{{\mathbb M}}})^{-1/2}\det_{2[\perp{\bf\hat k}],
\text{rep}}\left[{\sf Q}_{\bf\hat k}.\left({\bf\hat k}.
\overline{\overline{{\mathbb M}}}^{-1_{4s',\text{rep}}}.
{\bf\hat k}\right).{\sf Q}_{\bf\hat k}\right]^{-1/2},
\end{eqnarray}
\end{mathletters}
and ${\sf Q}_{\bf\hat k}={\sf I}-{\bf\hat k}{\bf\hat k}$, 
and ${\bf\hat k}={\bf k}/k$. In the first step from (\ref{ligne1}) 
to (\ref{ligne2}), we used the formula (\ref{notinv2}) generalized 
to the replica space. The resulting factor 
$\delta(\mathop{\text{tr}}_2 {\bf k}\lambdabf)\propto
\delta({\bf\hat k}.\lambdabf)$ was absorbed in the measure 
$\mathop{d\lambdabf^\alpha}_{[\perp{\bf\hat k}]}$ defined 
in (\ref{vecperp}). Next, we appealed to (\ref{lamperp}), 
also extended to replica space. Equ.\ (\ref{znav}) follows. 
Had $\overline{\overline{\mathbb M}}$ been invertible in 
$L(4s,\text{rep})$, one would have found
\begin{equation}
\label{znavinv}
\left\langle Z^r\right\rangle\propto\langle 
e^{-r\beta {v\over V} \psi_{\bf x}(\Sigmasf)}
\rangle^{V\over v}\prod_{\bf k\not={\bf 0}}
\left\{
\det_{4s,\text{rep}}(\overline{\overline{{\mathbb M}}})
\det_{2,\text{rep}}\left[\left({\bf\hat k}.
\overline{\overline{{\mathbb M}}}^{-1_{4s,\text{rep}}}.
{\bf\hat k}\right)\right]
\right\}^{-1/2}
\end{equation}
instead.

%-----------------------------------------------------------------------------
\section{The Legendre transform}
\label{tlt}
Though the plastic dissipation potential (\ref{nortplast}) and the 
viscoplastic potential (\ref{nortvis}) have been obtained for 
different boundary conditions, we show here that they are linked 
by the Legendre transform (\ref{legendre}). 

Let us deduce (\ref{nortplast}) from (\ref{nortvis}), for instance, 
assuming that (\ref{legendre}) holds between both. We write 
$\Phi({\sf D})=\Phi_0({\sf D})+\delta\Phi({\sf D})$, $\Psi(\Sigmasf)
=\Psi_0(\Sigmasf)+\delta\Psi(\Sigmasf)$, where $\Phi_0({\sf D})
=\langle\phi_{\bf x}({\sf D})\rangle$, $\Psi_0(\Sigmasf)=\langle
\psi_{\bf x}(\Sigmasf)\rangle$ are the leading terms in the 
perturbative expansions, and $\delta\Phi({\sf D})$ and 
$\delta\Psi(\Sigmasf)$ are the corrective terms.  It is straightforward 
to check that $\Phi_0({\sf D})$ and $\Psi_0(\Sigmasf)$ are 
Legendre duals: $\Phi_0({\sf D})+\Psi_0(\Sigmasf)=\Sigmasf
\!:\!{\sf D}$. Hence
\begin{equation}
\Phi({\sf D})=\Sigmasf\!:\!{\sf D}-\Psi_0(\Sigmasf)-
\delta\Psi(\Sigmasf)=\Phi_0({\sf D})-\delta\Psi(\Sigmasf),
\end{equation}
so that $\delta\Phi({\sf D})=-\delta\Psi(\Sigmasf)$. Moreover, 
since $\Sigmasf={\partial \Phi({\sf D})/\partial{\sf D}}={\partial 
\Phi_0({\sf D})/\partial{\sf D}}+{\partial \delta\Phi({\sf D})/
\partial{\sf D}}$, we obtain to second order
\begin{equation}
\label{pertphi}
\Phi({\sf D})\simeq\Phi_0({\sf D})-\delta\Psi\bigl({\partial 
\Phi_0({\sf D})/\partial{\sf D}}\bigr).
\end{equation}
The Legendre transform of $\Psi_0(\Sigmasf)=\langle\omega_n
\rangle\Sigma_{\text{eq}}^{n+1}/(n+1)$ reads, with $m=1/n$:  
\begin{equation}
\Phi_0({\sf D})=\langle \omega_n\rangle^{-m}
D_{\text{eq}}^{m+1}/(m+1), \quad \mathop{\text{tr}}_2({\sf D})=0.
\end{equation}
In addition,  
\begin{equation}
\Sigmasf'\simeq {d-1\over d}\langle \omega_n\rangle^{-m}
D_{\text{eq}}^{m-1}{\sf D}, \qquad\Sigma_{\text{eq}}\simeq
\langle \omega_n\rangle^{-m}D_{\text{eq}}^{m}.
\end{equation}
These expressions for $\Sigmasf'$ and $\Sigma_{\text{eq}}$ 
are correct to first order only. However, symmetry considerations 
show that $\Sigmasf'$ is always proportional to ${\sf D}$. 
Thus ${\Sigmasfhat}={\sf\hat D}$, whence from the 
definitions (\ref{dhatk}) and (\ref{sighatk}),
\begin{equation}
\Sigma_{\bf\hat k}=1-2 D_{\bf\hat k}.
\end{equation}

The weak-disorder expansion of $\Phi({\sf D})$ has to be 
expressed in terms of $\theta_m=\omega_n^{-m}=\langle 
\theta_m\rangle+\delta\theta_m$. To second order, we have
\begin{equation}
\langle\omega_n\rangle\simeq\langle\theta_m\rangle^{-1/m}
\left[1+{1\over 2}{m+1\over m^2}{\langle\delta\theta_m^2
\rangle\over \langle\theta_m\rangle^2}\right], \qquad 
{\left\langle\delta\omega_n^2\right\rangle\over 
\left\langle\omega_n\right\rangle^2}\simeq{1\over m^2}
{\left\langle\delta\theta_m^2\right\rangle\over 
\langle\theta_m\rangle^2}.
\end{equation}
Gathering these results into the perturbative expansion 
(\ref{pertphi}) of $\Phi({\sf D})$ finally yields 
(\ref{nortplast}) back, as announced.

%-----------------------------------------------------------------------------

\end{document}